\documentclass[11pt,a4paper]{article}
\usepackage{jheppub}
\usepackage[latin9]{inputenc}
\usepackage{amsmath}
\usepackage{amssymb}
\usepackage{setspace}
\usepackage{graphicx}
\usepackage{babel}

\title{Quasicrystals in QCD}

\author[a,b]{Zebin Qiu\,
}
\emailAdd{qiuzebin@nt.phys.s.u-tokyo.ac.jp}

\author[a,b,c]{and Muneto Nitta\,
}
\emailAdd{nitta@phys-h.keio.ac.jp}
\affiliation[a]{%
Research and Education Center for Natural Sciences, Keio University, 4-1-1 Hiyoshi, Yokohama, Kanagawa 223-8521, Japan}
\affiliation[b]{
Department of Physics, Keio University, 4-1-1 Hiyoshi, Yokohama, Kanagawa 223-8521, Japan}
\affiliation[c]{
International Institute for Sustainability with Knotted Chiral Meta Matter(SKCM$^2$), Hiroshima University, 1-3-2 Kagamiyama, Higashi-Hiroshima, Hiroshima 739-8511, Japan
}

\abstract{We study the ground state of the low energy dense QCD  with the assumption of chiral condensates of quarks.
Under an external magnetic field, mesons could form soliton lattices via the chiral anomaly.   
For such scenarios, we present a unified description of pions and $\eta$ meson with a $U(2)$ field in the framework of the chiral perturbation theory. 
Our result shows the ground state is a mixture of the magnetized domain walls formed by neutral pion $\pi^0$ and $\eta$ meson when they coexist. 
The winding number of the ground state would alter according to the strength of the magnetic field.
When the magnetic field is strong or the chemical potential is large, the proportion of the mixture is determined by the decay constants and the contributions to the anomalous action of $\pi^0$ and $\eta$ meson. The resulting configuration is either a mixed soliton lattice or a quasicrystal 
which could be dubbed a ``chiral soliton quasicrystal''.
}

\begin{document}

\maketitle

\section{Introduction}

The phase structure of QCD is of tremendous research interest for
decades. Especially, under external fields such as magnetic field
or rotation, richer phase structures with nontrivial topology arise
due to the coupling of these external fields with the baryon chemical potential
via the chiral anomaly.
Recently,
there has been a challenging new idea that the ground state of dense
QCD matter in the presence of a strong magnetic field could be the chiral
soliton lattice (CSL) made of bosonic neutral pion $\pi^{0}$ instead
of fermionic baryons~\cite{Son:2007ny,Eto:2012qd,Brauner:2016pko,Brauner:2019aid,Brauner:2021sci}.
Following this lead, related phases involving charged pions
have then been studied, including the charged pion condensate \cite{Brauner:2016pko}  
and its competition with
an Abrikosov vortex lattice \cite{Evans:2022hwr}, for which one can draw the phase diagram
parameterized by the magnetic field and the isospin chemical potential~\cite{Gronli:2022cri}.
Further studies of CSLs are devoted to 
thermal fluctuations  \cite{Brauner:2017uiu,Brauner:2017mui,Brauner:2021sci,Brauner:2023ort},
rapid rotation 
\cite{Huang:2017pqe,Nishimura:2020odq,Eto:2021gyy,Chen:2021aiq}, 
and quantum nucleation  
\cite{Eto:2022lhu,Higaki:2022gnw}, 
see also Refs.~\cite{Yamada:2021jhy,Brauner:2019rjg} for related issues.

These novel phenomena could all be well described by the chiral perturbation
theory (ChPT) which is a low-energy effective theory of QCD.
It is based on the chiral symmetry breaking $U(1)_{V}\times SU(N_{f})_{L}\times SU(N_{f})_{R}\rightarrow U(1)_{V}\times SU(N_{f})_{V}$
that leaves bound states of quarks as the Nambu-Goldstone (NG) bosons, specifically pions.
The effective action of ChPT is therefore in terms of the $SU(N_{f})$
field, constructed by a perturbative expansion with a power counting
scheme on momentum. Effects from chiral anomaly are encoded in the
Wess-Zumino-Witten (WZW) term \cite{Son:2004tq,Son:2007ny}. Describing physics on the same scales of
density and magnetic field, 
the simplest extension of the ChPT could
be the addition of a gauged Skyrme term which stabilises a soliton
solution describing the baryon. In such a framework, the aforementioned
pionic phases could be compared with a multi-baryon crystal called
Skryme crystal under a magnetic field. Latest efforts have found the
CSL could be taken over by the magnetized Skyrme crystal in a realm of higher density and/or lower magnetic field among the phase diagram~\cite{Chen:2021vou,Chen:2023jbq} 
(see also Ref.~\cite{Kawaguchi:2018fpi}). 
More recently, 
the higher density region of the CSL phase 
has been found to transit to a domain-wall Skyrmion phase,
in which a Skyrmion is decomposed as a baby skyrmion accommodated in a pion domain wall~\cite{Eto:2023lyo}, attracting its own
research interests in the sense of the topology~\cite{Nitta:2012wi,Nitta:2022ahj,Ross:2022vsa}.

So far the phases mentioned above are centred on pions as building
blocks. But there could be a similar anomalous coupling of ``heavier''
NG bosons. In the baryon density window next to (higher than) that
of the CSL, there comes the $\eta$ meson, resulting
from the $U(1)_{A}$ axial symmetry breaking. Strictly, the 
$U(1)_{A}$ symmetry is not a symmetry of QCD but is violated by anomalous
fluctuations of gluons. However at the large density where the color
Deybe screening applies and instanton effects are suppressed, the
$U(1)_{A}$ symmetry could be treated as a symmetry approximately.
Its spontaneous breaking gives rise to the $\eta$ meson whose mass
is heavier than pions yet still light compared to the chemical potential.
In such a density regime, the $\eta$ meson could couple with a magnetic field and form a similar axial domain wall as $\pi^{0}$~\cite{Son:2004tq, Son:2007ny}.
Interestingly, while the $\pi^{0}$ pion does not couple with a rotation,
the $\eta$ meson does. The result is a rotational counterpart of the $\pi^{0}$ CSL, 
the $\eta$ CSL~\cite{Huang:2017pqe, Nishimura:2020odq,Eto:2021gyy}.
Natural playgrounds to study anomalous effects of $\eta$ mesons are
the 2-flavor superconducting (2SC)~\cite{Alford:1997zt, Rapp:1997zu}
and the color-flavor locked (CFL)~\cite{Alford:1998mk} phases.

For the physics we would like to discuss in the present paper, $N_{f}=2$
(2-flavor) case suffices. We are interested in the effect of a magnetic
field that couples with both the pions and the $\eta$ meson. Though usually these particles are considered at separate density scales, in principle
they can coexist in nuclear matter. Lately, there have been attempts to put them in the same theoretical framework of the ChPT~\cite{Aoki:2014moa,Eto:2021gyy}, where the field is augmented to $U(2)=[U(1)\times SU(2)]/\mathbb{Z}_{2}$,
encompassing the $U(1)$ of $\eta$ and the $SU(2)$
of pions. 
In such a scenario, more intriguing phase structures appear.
It was pointed out in Ref.~\cite{Aoki:2014moa} that one can realize the $CP$-violating Dashen phase by tuning the masses of $u,d$ quarks. 
In that paper, neither external fields nor interactions of pions and $\eta$ are considered.
It is important for us to understand that 
in the $U(2)$ ChPT
even mass terms alone could yield nontrivial solutions other than vacuum. 
Furthermore, in Ref.~\cite{Eto:2021gyy} the coupling of the $\eta$ meson to a rotation
is included. 
Then, depending on the magnitude of the rotation and coupling constants, there exist three possible ground state phases, $\eta$ CSL and two kinds of non-Abelian 
CSLs, called ``dimer'' and ``deconfined'' phases. 
Here, the term ``non-Abelian'' implies that 
a soliton has non-Abelian moduli coming from 
spontaneous symmetry breaking of the vacuum symmetry 
in the vicinity of the soliton.
When the sine-Gordon soliton is embedded in $SU(N)$, 
it is a non-Abelian soliton with ${\mathbb C}P^{N-1}$ moduli \cite{Nitta:2014rxa,Eto:2015uqa}.
Such a phenomenon is one direct inspiration for the present work. 

Crystalline configurations with periodicity 
have been widely studied 
in the context of QCD; apart from 
chiral soliton lattices which we focus on in this paper, 
extensive investigations have been conducted on modulated phases in Nambu-Jona Lasino 
or Gross-Neveu models 
\cite{Nakano:2004cd,
Nickel:2009ke,Basar:2009fg,
Buballa:2014tba,
Hidaka:2015xza}, 
crystalline color supercondutors \cite{Casalbuoni:2003wh,Anglani:2013gfu}, 
and so on.
On the other hand,    
in condensed matter materials, there are many examples of 
{\it quasicrystals}.
Quasicrystals are aperiodic incommensurate lattice
structures 
\cite{Steinhardt,JANSSEN198855,DiVincenzo,Janot,Jansen:2007,Stadnik,Fan,Scott,Jaric}. 
See Refs.~\cite{Baggioli:2020haa,Surowka:2021ved} for 
recent studies of quasicrystals based on field theory.

In this paper, we make the first step of the study on quasicrystal in QCD, 
by considering the effect of an external magnetic field 
on the $U(2)$ NG fields 
in the chiral symmetry breaking phase, 
particularly the $\pi^{0}$ pion and $\eta$ meson.
Our scenario with an external magnetic field brings an essential difference
in the anomalous action compared to that of a rotation. 
The anomalous action receives contributions from both $\pi^{0}$ pion and $\eta$ meson, 
and the resulting magnetic
energy determines the ground state.
We witness that pure $\pi^0$ solitons often have the lowest energy (being ground states) and
a pure $\eta$ soliton always has much higher energy. 
Nevertheless, somewhat surprisingly, the ground state can also be 
 a ``mixed'' soliton lattice
with the nontrivial winding of both mesons, 
rather than a separate $\pi^{0}$
or $\eta$ CSL. 
More generally, the ground state could even be a one-dimensional 
quasicrystal along the direction of the 
magnetic field. 
In the strong magnetic field and/or 
large chemical potential, 
we find that the ratio of the decay constants 
of pions and $\eta$ meson plays a crucial role.
Specifically, when the ratio happens to be rational, 
we would have a lattice or crystal. Otherwise,
we would have a quasicrystal. 
To the best of our knowledge, 
this is the first quasicrystal in 
a realistic setup of QCD.

This paper is organized as follows.
In Sec.~\ref{sec:U(2)}, our setup 
of the  $U(2)$ ChPT featuring pions and $\eta$ meson is given.
In sec.~\ref{sec:mixed}, 
the simplest mixed soliton lattice is investigated 
in comparison to pure $\pi_0$ and $\eta$ CSLs.
In Sec.~\ref{sec:GS}, 
the ground states varying with the magnetic field are determined 
assuming a periodicity 
up to a certain soliton number per lattice period.
In Sec.~\ref{sec:strong-B}, 
The limit of a strong magnetic field and/or large chemical potential is discussed, where the quasicrystal emerges.
Sec.~\ref{sec:summary} is devoted to a conclusive summary and outlook discussions. 
Through this work we adopt  $g^{\mu\nu}=\left(+,-,-,-\right)$.

\section{$U(2)$ Chiral Perturbation Theory in a Magnetic Field}\label{sec:U(2)}

We consider non-Abelian chiral soliton lattices constructed by the
two-flavor $U(2)$ field: 
\begin{equation}
U=\exp\left(i\eta/f_{\eta}\right)\Sigma;\quad\Sigma=\exp\left(i\tau^{k}\pi_{k}/f_{\pi}\right);\quad k=1,2,3.\label{eq:U2}
\end{equation}
The $f_{\eta}$ and $f_{\pi}$ are the decay constants of 
the eta meson $\eta$
and pions $\pi^{k}$, respectively. It suffices to adopt the low-energy effective
theory based on the chiral symmetry breaking for our purpose.

The chiral Lagrangian featuring the kinetic and the mass terms reads:
\begin{align}
\mathcal{L}_{\text{chiral}}= & \frac{1}{2}\partial_{\nu}\eta\partial^{\nu}\eta+\frac{f_{\pi}^{2}}{4}\mathrm{Tr}\left(\partial_{\mu}\Sigma\partial^{\mu}\Sigma^{\dagger}\right)\nonumber \\
 & +\left\{ \frac{a}{2}\left(\det U-1\right)+b\,\text{Tr}\left[M\left(U-1\right)\right]+\text{h.c.}\right\} .\label{eq:Lchiral}
\end{align}
Here, $b$ is a constant associated with the usual mass matrix $M$. 
$M$ is taken to be simply $M=m\mathbb{I}$ because we consider the regime
with chemical potential much larger than quark masses so the masses
of $u,d$ quarks can be regarded as approximately the same $m$. On
top of that, $a$ represents the extra mass given to the $\eta$ meson.
We have quoted the form of $\mathcal{L}_{\text{chiral}}$ from
Ref.~\cite{Eto:2021gyy} which studied thoroughly, within the same
framework, the anomalous coupling of 
the $U(2)$ field
to a rotation.
In the present work, differently, we consider an external magnetic
field instead of the rotation. 

It is well known that charged pions $\pi_1, \pi_2$
become very ``massive'' under a strong magnetic field while the neutral pion 
$\pi_{3}$ remains the only NG boson among the pions.
Focusing on the ground state properties, we ignore charged pions and
set $\pi_{1,2}=0$ throughout this work.

Whilst rotation couples to
the $\eta$ meson only, a magnetic field couples with both the $\eta$ meson and neutral pion $\pi_{3}$ 
via the axial anomaly, contributing an extra term in Lagrangian~\cite{Son:2004tq,Son:2007ny}:
\begin{equation}
\mathcal{L}_{B}=\frac{1}{8\pi^{2}}\epsilon^{\mu\nu\alpha\beta}\sum_{i=0,3}C_{i}\partial_{\mu}\phi_{i}A_{\nu}^{\text{B}}F_{\alpha\beta},\label{eq:Lanom}
\end{equation}
where the $\phi_{i}$ with $i=0,3$ denote relevant NG bosons respectively,
i.e., $\eta$, $\pi_{3}$ in our case. In other words, $\phi_{i}$
are the dimensionless 
\begin{equation}
\phi_{0}\equiv\frac{\eta}{f_{\eta}},\quad\phi_{3}\equiv\frac{\pi_{3}}{f_{\pi}},
\label{eq:phi}
\end{equation}
appearing in the exponentials in Eq.~\eqref{eq:U2}. The $F_{\alpha\beta}$
is the electromagnetic field strength. Without loss of generality,
we set up a homogeneous background magnetic field along longitudinal
direction $\hat{z}$, specifying $F_{xy}=-F_{yx}=-B$. The $A_{\nu}^{\text{B}}=\left(\mu,\boldsymbol{0}\right)$
is a common choice of baryon gauge field to feature the chemical potential
$\mu$. Then importantly, Eq.~\eqref{eq:Lanom} $C_{i}$ are the anomalous
coefficients signifying the interplay between $U(1)_{\text{EM}}$
and $U(1)_{\text{B}}$ gauge fields via 
the chiral anomaly:
\begin{equation}
C_{i}=\sum_{a=u,d}Q_{5,i}^{a}Q_{i}^{a},\label{eq:Canom}
\end{equation}
in which $Q_{i}^{a}$ and $Q_{5,i}^{a}$ are charges and axial charges
of quarks comprising the relevant NG bosons labeled by index $i$.
In our 2-flavor model, it certainly reads
\begin{align}
\eta\text{: } & Q_{5}^{u}=Q_{5}^{d}=1,\\
\pi_{3}\text{: } & Q_{5}^{u}=-Q_{5}^{d}=1.
\end{align}
Needless to say $Q^{u}=2/3$ and $Q^{d}=-1/3$. With such combinations,
we finally arrive at the anomalous Lagrangian in our setup: 
\begin{equation}
\mathcal{L}_{B}=\frac{\mu}{4\pi^{2}}\boldsymbol{B}\cdot\left(\boldsymbol{\nabla}\phi_{3}+\frac{1}{3}\boldsymbol{\nabla}\phi_{0}\right),\label{eq:LB}
\end{equation}
which should be added to Eq.~\eqref{eq:Lchiral} to complete our effective
theory
\begin{equation}
\mathcal{L}=\mathcal{L}_{\text{chiral}}\big|_{\pi_{1,2}=0}+\mathcal{L}_{B}.
\end{equation}

We remark that usually the $\eta$ meson becomes relevant in the density
regime above the low-density nuclear matter, 
e.g., the 2SC or CFL phase. For
sufficiently large $\mu\gg\Lambda_{\text{QCD}}$, both $f_{\eta}$
and $f_{\pi}$ can be evaluated in terms of $\mu$. However, in this
work, we regard them as independent and tunable parameters.

As indicated by the established studies on the CSL lattices,
derivative terms involving transverse directions $\partial_{x,y}$
should vanish in the ground state because they contribute to the soliton energy functional in a positive-definite way. The same argument applies
to our model when considering soliton solution with null $\partial_{0}$.
Therefore the issue we tackle is essentially 1D, depending on the
longitudinal coordinate $z$ selected by $\boldsymbol{B}=B\hat{z}$.
The equation of motion (EOM) is equivalent to the variational principle
minimizing the Hamiltonian or energy functional $H$ corresponding
to $\mathcal{L}$. To be specific, we write the energy functional:
\begin{align}
\tilde{H}= & \frac{1}{2}\left(\alpha\phi_{3}^{\prime2}+\phi_{0}^{\prime2}\right)-\frac{\gamma}{2\pi}\left(\phi_{3}^{\prime}+\frac{1}{3}\phi_{0}^{\prime}\right)\nonumber \\
 & +\sin\beta\left(1-\cos2\phi_{0}\right)+\cos\beta\left(1-\cos\phi_{0}\cos\phi_{3}\right),\label{eq:H}
\end{align}
where we have employed dimensionless quantities similar to those in Ref.~\cite{Eto:2021gyy}:
\begin{equation}
\tilde{z}\equiv\frac{\left[a^{2}+\left(4mb\right)^{2}\right]^{1/4}}{f_{\eta}}z,\quad\tilde{H}=\frac{H}{\sqrt{a^{2}+\left(4mb\right)^{2}}},\label{eq:rescale}
\end{equation}
with $\prime$ implying 
$\partial/\partial\tilde{z}$. 
With the help
of Eq.~\eqref{eq:rescale}, the parameters $a,b$ are rephrased into 
\begin{equation}
\alpha\equiv\frac{f_{\pi}^{2}}{f_{\eta}^{2}},\quad\beta\equiv\arctan\frac{a}{4mb},
\end{equation}
and the magnetic field $B$ is encoded in 
\begin{equation}
\gamma\equiv\frac{\mu B}{2\pi}\frac{\left[a^{2}+\left(4mb\right)^{2}\right]^{-1/4}}{f_{\eta}}.
\end{equation}
In what follows, we abbreviate the tildes on top of $H$ and $z$ and
deal with above-mentioned dimensionless quantities by default. The EOM is derived from the variational principle
exerted on $H$:
\begin{align}
\alpha\phi_{3}^{\prime\prime} & =\cos\beta\cos\phi_{0}\sin\phi_{3}\nonumber \\
\phi_{0}^{\prime\prime} & =\cos\beta\cos\phi_{3}\sin\phi_{0}
+2\sin\beta\sin2\phi_{0}.\label{eq:eom}
\end{align}
Obviously if one of $\phi_{0}$ and $\phi_{3}$ is absent, the EOM
reduces simply to the sine-Gordon equation whose solution is analytically
known. In our work, the highlight is the mixture of $\phi_{0}$ and
$\phi_{3}$ obeying a coupled sine-Gordon equation (\ref{eq:eom}).

\section{Mixed Soliton Lattice}\label{sec:mixed}

In this section, we start with the simplest mixed soliton lattice solution
in our model with assuming a periodicity. 
In general, we intend to solve the $U\in [U(1)\times U(1)]/\mathbb{Z}_{2}$
field over the distance $d$ along the $z$-direction. The winding is
depicted by the boundary condition:
\begin{equation}
\left(\phi_{3},\phi_{0}\right)\bigg|_{z=0}=\left(0,0\right),\quad\left(\phi_{3},\phi_{0}\right)\bigg|_{z=d}=\left(p\pi,q\pi\right),\label{eq:bcpq}
\end{equation}
where the single-value condition $\Sigma\left(d\right)=\Sigma\left(0\right)$
restricts the choice of $p$ and $q$:
\begin{equation}
\frac{p\pm q}{2}\in\mathbb{Z}.\label{eq:singlevalue}
\end{equation}
The familiar winding $\pi_{1}[U(1)]\simeq \mathbb{Z}$
applies to each separate $U(1)$ in $\Sigma$. The corresponding
topological charges are $\pi^{0}$ and $\eta^{\prime}$ numbers respectively. In our framework, we prove numerically that 
they are reproduced by the
following boundary conditions:
\begin{align}
\text{\ensuremath{\pi_{0}}\text{: }} & \left(p,q\right)=\left(2,0\right),\label{eq:20}\\
\eta\text{: } & \left(p,q\right)=\left(0,2\right).\label{eq:02}
\end{align}
In comparison, we innovatively introduce the concept of a ``mixed''
soliton with both $p,q\neq0$. The simplest case abiding by Eq.~\eqref{eq:singlevalue}
is
\begin{equation}
\text{(Simplest) Mixed Soliton: }\left(p,q\right)=\left(1,1\right).\label{eq:11}
\end{equation}

Before proceeding, to give a better perception of the mixed soliton,
let us mention a special case of $\beta=\pi/2$, which is analytically
solvable. The EOMs ~\eqref{eq:eom} reduce to 
\begin{equation}
\alpha\phi_{3}^{\prime\prime}=0,\quad\phi_{0}^{\prime\prime}=2\sin2\phi_{0},
\end{equation}
whose solution reads
\begin{equation}
\phi_{3}=\frac{p\pi z}{d},\quad\phi_{0}=\text{am}\left(\nu z,-\frac{4}{\nu^{2}}\right),\label{eq:ana}
\end{equation}
where am is the Jacobi amplitude and $\nu$ satisfies $\phi_{0}\left(d\right)=q\pi$,
dictated by the boundary conditions \eqref{eq:bcpq}. Such a solution,
with nonzero $p$ and $q$, exemplifies the mixed soliton solution.

Now we come to the case with general $\alpha$ and $\beta$, the EOMs
are solved numerically over distance $d$. We apply the solutions
into $H$, evaluating the soliton energy. Notably, in the context
of a soliton lattice, $d$ can be regarded as the lattice period.
For a fixed total lattice size, we need to find the $d$ that minimizes
the total lattice energy density
\begin{equation}
\mathcal{E}=\frac{1}{d}\int dzH.
\end{equation}
If $\mathcal{E}$ has a global minimum, $d_{L}=\text{argmin}\mathcal{E}\left(d\right)$
would be the lattice period of the ground state. Whereas $\mathcal{E}$
as a function of $d$ does not always feature a minimum. It depends
on the value of the magnetic field, or in other words, $\gamma$. There
should be the critical value of $\gamma$ only above which can $d_{L}$
exist. In this section, we discuss the critical $\gamma$ physics by
limiting our attention to the three building block configurations
Eqs.~\eqref{eq:20}, \eqref{eq:02}, and \eqref{eq:11}. Quantities
associated with each configuration would be dressed by lower index
``$\pi$'', ``$\eta$'' and ``m'', respectively. We observe for all
three configurations, there are critical $\gamma_{\pi,\eta,\text{m}}$
for $\mathcal{E}_{\pi,\eta,\text{m}}$ to exhibit a global minimum.
\begin{equation}
\text{For }\gamma>\gamma_{\pi,\eta,\text{m}},\quad\exists d_{L}\in\left(0,d\right):\quad\mathcal{E}_{\pi,\eta,\text{m}}^{\prime}\left(d_{L}\right)=0.
\end{equation}
$\gamma_{\pi,\eta,\text{m}}$ are located numerically and would be detailed
in what follows.

\begin{figure}      
\centering       
\includegraphics[width=0.75\columnwidth]{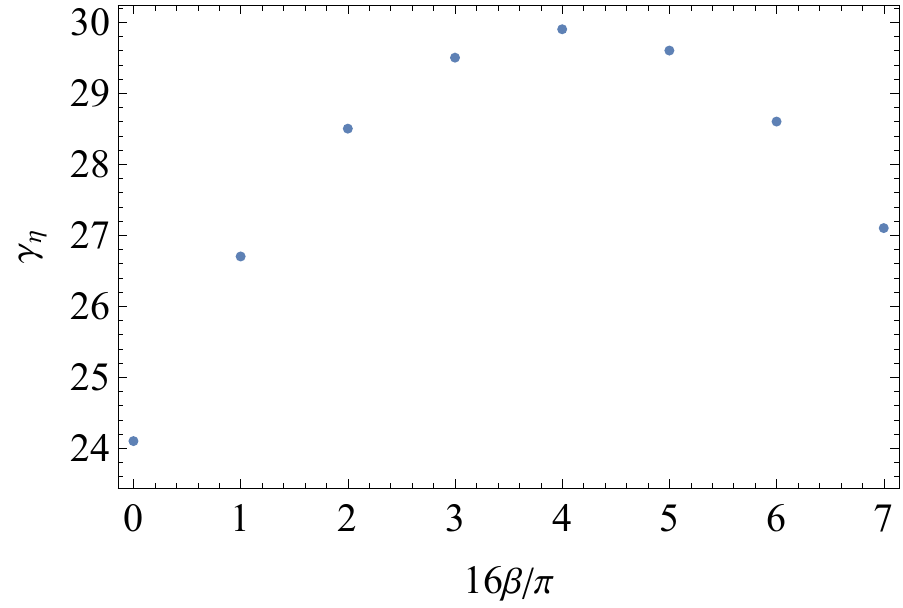}      
\caption{Critical $\gamma$ for $\eta^\prime$ soliton lattice}     
\label{fig:gamma2}  
\end{figure}

First, we find $\gamma_{\eta}$ is independent of $\alpha$. This
is easy to understand because the only field involving $\alpha$ in $H$ is the $\phi_{3}$, and it proves numerically that $\phi_{3}\equiv0$
for $\left(p,q\right)=\left(0,2\right)$. We therefore display $\gamma_{\eta}$ under several
different values of $\beta\in\left(0,\pi/2\right)$ in Fig.~\ref{fig:gamma2} , which is qualified
as a summary of our results about the separate $\eta$ soliton lattice
under magnetic field. We shall remind the $\beta=\pi/2$ case is
excluded since it has already been solved analytically by Eq.~\eqref{eq:ana}.

Next, we shall report the observation that $\gamma_{\eta}\sim\mathcal{O}\left(3\gamma_{\pi}\right)$
which is reminiscent of the discrepancy of anomalous coefficients in 
Eq.~\eqref{eq:Canom} by the prefactor $1/3$. More explicitly, this
can be seen from the topological part, 
Eq.~(\ref{eq:LB}), 
among $\mathcal{E}$, depending
on the $\gamma$, i.e.,
\[
\mathcal{E}_{\text{topo}}=-\frac{\gamma}{d}\cdot\begin{cases}
1 & \text{for } \pi^{0}\\
1/3 & \text{for } \eta\\
2/3 & \text{for ``mixed"}
\end{cases}.
\]
In other words, comparing the two separate soliton lattices under
a common magnetic field, we can naturally expect $\eta$ CSL to
bear higher energy than $\pi^{0}$-lattices. This explains why the magnetic effect of the $\eta$ soliton lattices is often
overlooked in lower density/energy regimes; the $\pi^{0}$-lattice
demands a lower critical magnetic field to emerge as the ground state.
Along this line, one might intuitively guess that the mixed soliton
lattice, as a mixture of $\pi^{0}$ and $\eta$ soliton lattices,
would feature intermediate energy. But astonishingly we find that it
is not the case. A mixed soliton lattice can have the lowest energy in a certain parameter region!

\begin{figure}      
\centering       
\includegraphics[width=0.75\columnwidth]{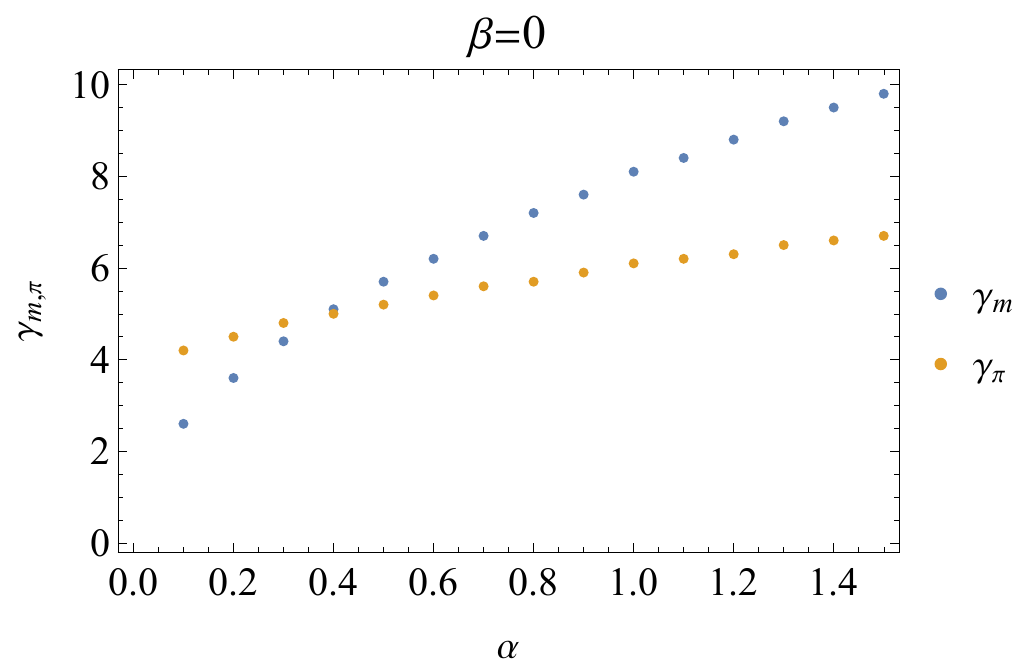}      
\caption{Critical $\gamma_{\text{m},\pi}$ at $\beta=0$ and varied $\alpha$.}     
\label{fig:gamma13beta0}  
\end{figure}

\begin{figure}      
\centering       
\includegraphics[width=0.75\columnwidth]{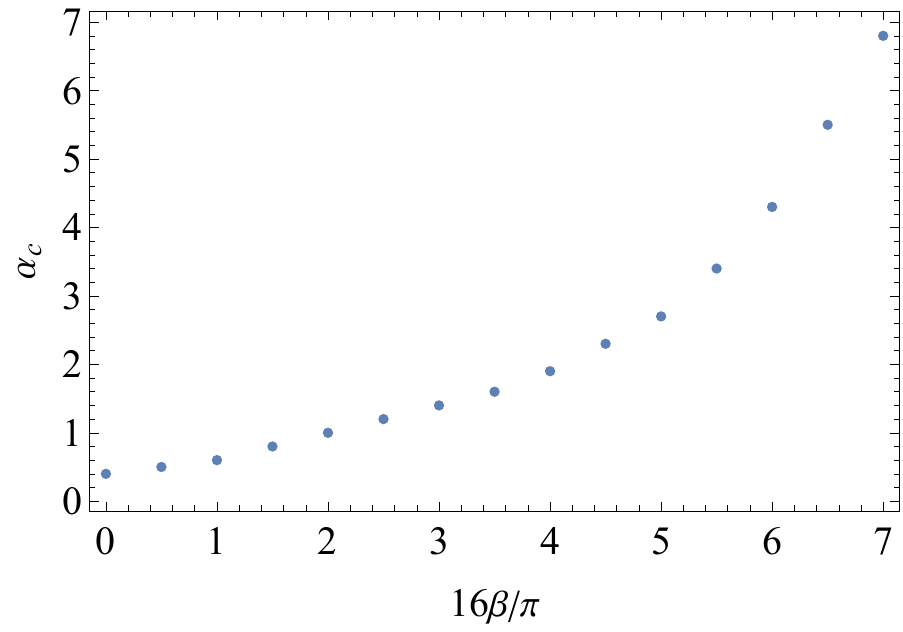}      
\caption{Critical $(\alpha_c,\beta)$ for the ground state to be the mixed soliton lattice.
The dots split the $\alpha$-$\beta$ plane into two regions. 
In the upper region above the $\alpha_{c}\left(\beta\right)$, we have $\gamma_{\text{m}}<\gamma_{\pi}$.}     
\label{fig:ab}  
\end{figure}

Ultimately we come to address such a duel between $\gamma_{\text{m}}$
and $\gamma_{\pi}$. Which one is lower depends on parameters $\alpha$
and $\beta$. 
We illustrate the dependence of $\gamma_{\pi,\text{m}}$
on $\alpha$ in the case with $\beta=0$, as shown in Fig.~\ref{fig:gamma13beta0}.
Importantly, for $\alpha>\alpha_{c}=0.4$, one can witness $\gamma_{\text{m}}<\gamma_{\pi}$
and during this window $\gamma\in\left(\gamma_{\text{m}},\gamma_{\pi}\right)$
the mixed soliton lattice arises while neither does the $\pi^{0}$ CSL nor
the $\eta$ CSL. 
In general, $\alpha_{c}=\alpha_{c}\left(\beta\right)$ depends on
$\beta$, which is displayed in Fig.~\ref{fig:ab}. Therein, for the
parameter region above the $\alpha_{c}\left(\beta\right)$ line, we
conclude with $\gamma_{\text{m}}<\gamma_{\pi}$ that the mixed soliton
lattice is the ground state that demands a lower critical magnetic field and features a lower energy, in comparison to the stand-alone
$\pi^{0}$ or $\eta$ CSL.

Nevertheless, this is not the end of the story, first, there can be
other (actually numerous) modes of mixing solitons, as dictated by the $p,q$
in Eq.\eqref{eq:bcpq}. 
The one demonstrated above is only the simplest
one with $p=q=1$. Moreover, even in the parameter region where
$\gamma_{\text{m}}<\gamma_{\pi}$, if we tune the magnetic field to
$\gamma>\gamma_{\pi}>\gamma_{\text{m}}$, we find that $\pi^{0}$ CSL could win the competition against the mixed soliton lattice and take up the ground state, depending on the specific value of $\gamma$. We will
further explain these points in the next section.
Before that, it is
worth thinking about the mechanism that makes mixed soliton lattice energetically
favorable, for relatively small $\beta$ and large $\alpha$ according
to Fig.~\ref{fig:ab}. The favor of large $\alpha$ for a mixed
soliton lattice is explained by the kinetic term of Hamiltonian $\tilde{H}_{\text{kin}}=\frac{\alpha}{2}\phi_{3}^{\prime2}+\frac{1}{2}\phi_{0}^{\prime2}$
because nonzero variation of $\phi_{0}$ will take the load off $\phi_{3}^{\prime}$
if the weight $\alpha$ is heavy. The leverage of small $\beta$ can
be seen from the mixed mass term in the Lagrangian $\tilde{H}_{\text{mix}}=\cos\beta\left(1-\cos\phi_{0}\cos\phi_{3}\right)$
which could be lower for $\text{\ensuremath{\phi_{0,3}}}\in\left(0,\pi\right)$
than that for $\phi_{0}=0$ or $\phi_{3}\in\left(0,2\pi\right)$.
In this way, we have figured out the three basic lattice configurations
with the $\pi^{0}$, $\eta$, and the mixed soliton of $\left(p,q\right)=\left(1,1\right)$.
Later on, configurations of higher $p,q$ can be regarded as a further
mixture using these three ingredients with different proportions.

\section{Ground State Alternation}\label{sec:GS}

After figuring out critical $\gamma_{\text{m},\pi,\eta}$, we readily
discuss the phases of ground state under a varying $\gamma$. Definitely we are more interested in the parameter region above the $\alpha_{c}\left(\beta\right)$-line
where a mixed soliton lattice can be the ground state with the lowest
$\gamma$. An exemplary case is with $\beta=\pi/16$ and $\alpha=0.7$ given that common phenomenology
results concur\footnote{Of course, the specific value depends on additional information of
flavor symmetry breaking and mixture of $\eta$ and $\eta^{\prime}$,
which are beyond our consideration and cannot be obtained in the
current framework. Thus we just choose an exemplary value. In contrast,
there are other choices from theoretical studies, e.g., in high density
limit (though strictly it is the 3-flavor CFL case which does not apply
here), one will have $f_{\eta}=0.87f_{\pi}$ instead.} on $f_{\eta}\simeq1.2f_{\pi}$.
In this case we locate $\gamma_{\text{m}}=6.5<\gamma_{\pi}=6.7$,
letting alone the much higher $\gamma_{\eta}=26.7$. It is confirmed
numerically that for the full range of the magnetic field, $\eta$-solitons always have the highest energy, compared to the others. Therefore
the $\eta$ soliton lattices can be taken out of our study for now.
However, the duel between the $\pi^{0}$ and the mixed soliton lattices
for the ground state brings us another magnetic field scale $\gamma_{\pi\text{m}}=7.2$,
which flips the sign of $\mathcal{E_{\pi}-\mathcal{E}_{\text{m}}}$.
That is to say, for $\gamma\in\left(\gamma_{\pi},\gamma_{\pi\text{m}}\right)$
both the $\pi^{0}$ and mixed soliton lattices arise, 
and the latter has
lower energy $\mathcal{E}_{\text{m}}<\mathcal{E}_{\pi}$, being the
ground state. While for $\gamma>\gamma_{\pi\text{m}}$, $\mathcal{E}_{\pi}<\mathcal{E}_{\text{m}}$
indicates the realm of the ground state is lost to the $\pi^{0}$-soliton
lattice. We call such a phenomenon the ground state alternation, among
different soliton lattice configurations.

Such ground state alternations can be generalized to more configurations
dictated by varied $\left(p,q\right)$. Actually, the three configurations
presented so far are those winding $U$ by $2\pi$, or $p+q=2$ in
the language of Eq.~\eqref{eq:singlevalue}. The next winding class
is naturally $p+q=4$. It yields two new independent configurations
$\left(p,q\right)=\left(1,3\right)$ and $\left(3,1\right)$, for
which we employed indices ``4'' and ``5'' to mark their physical
quantities respectively. Also, we clarify the general rule of naming
$\gamma_{\#*}$ is based on the connotation that $\mathcal{E}_{\#}-\mathcal{E}_{*}$
turns from positive to negative when $\gamma$ surpasses $\gamma_{\#*}$
from below. Now we are ready to present the booming magnetic field
scales when just two new configurations 4 and 5 are added into our
consideration. 
We then have the following sequences of inequalities of 
energies of various configurations:
\begin{align}
\gamma & \in\begin{cases}
\left[\gamma_{\text{m}},\,\gamma_{\pi}\right): & \left(p,q\right)=\left(1,1\right)\text{ only}\\
\left[\gamma_{\pi},\,\gamma_{\pi\text{m}}\right): & \mathcal{E}_{\text{m}}<\mathcal{E}_{\pi}\\
\left[\gamma_{\pi\text{m}},\,\gamma_{4}=7.3\right): & \mathcal{E}_{\pi}<\mathcal{E}_{\text{m}}\\
\left[\gamma_{4},\,\gamma_{4\text{m}}=8.8\right): & \mathcal{E}_{\pi}<\mathcal{E}_{\text{m}}<\mathcal{E}_{4}\\
\left[\gamma_{4\text{m}},\,\gamma_{5}=12.8\right): & \mathcal{E}_{\pi}<\mathcal{E}_{4}<\mathcal{E}_{\text{m}}\\
\left[\gamma_{5},\,\gamma_{4\pi}=13.0\right): & \mathcal{E}_{\pi}<\mathcal{E}_{4}<\mathcal{E}_{\text{m}}<\mathcal{E}_{5}\\
\left[\gamma_{4\pi},\,\gamma_{\eta}\right): & \mathcal{E}_{4}<\mathcal{E}_{\pi}<\mathcal{E}_{\text{m}}<\mathcal{E}_{5}\\
\left[\gamma_{\eta},\,\infty\right): & \mathcal{E}_{4}<\mathcal{E}_{\pi}<\mathcal{E}_{\text{m}}<\mathcal{E}_{5}<\mathcal{E}_{\eta}
\;.
\end{cases}
\label{eq:windows}
\end{align}
The leftmost energies after the colon are of the ground states, followed by those of several metastable states. 
Each transition is a first order transition.

If one only focuses on the ground state, the analysis can
be a bit simplified and the conclusion is: when $\gamma$ is increased
from zero, the ground state is taken up by lattices of $\left(p,q\right)=\left(1,1\right)$,
$\left(2,0\right)$ and $\left(3,1\right)$ sequentially:
\begin{align}
\text{Ground State }\left(p,q\right) & =\begin{cases}
\left(1,1\right) & \gamma\in\left[\gamma_{\text{m}},\,\gamma_{\pi\text{m}}\right)\\
\left(2,0\right) & \gamma\in\left[\gamma_{\pi\text{m}},\,\gamma_{4\pi}\right)\\
\left(3,1\right) & \gamma\in\left(\gamma_{4\pi},\,\infty\right) \;.
\end{cases}\label{eq:ground}
\end{align}
Actually the configuration $\left(3,1\right)$ (or $\left(1,3\right)$)
could be regarded as a further mixture between the simplest mixed
soliton and 
a pure $\pi^{0}$ or $\eta$. In such a way, one can imagine
infinite possibilities of merging different mixed solitons to form
different new types of lattices. The more possibilities one takes
into account, the more byzantine ground state alternations will emerge.
Certainly, we cannot exhaust all possible configurations of mixed
solitons but we do find a certain pattern of the ground state in
strong magnetic field and/or high density limit $\gamma\rightarrow\infty$,
which would be the topic of the subsequent section.

\begin{figure}      
\centering       
\includegraphics[width=0.75\columnwidth]{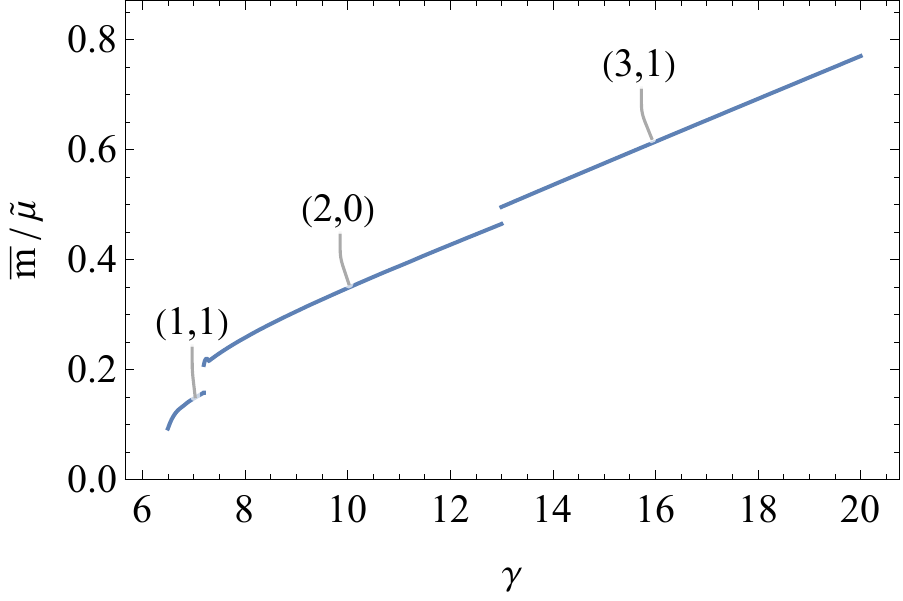}      
\caption{Magnetic moment density averaged over lattice period of alternating ground states.}     
\label{fig:staircase} 
\end{figure}

Before that, as a finale of the present section, we would quantify
the ground state alternation by the magnetic moment density (per unit
transverse area), which can be derived from its relation to the free
energy from
\begin{equation}
\mathrm{M}\equiv-\frac{\delta}{\delta B}\int dzH=-\frac{\delta}{\delta B}\left(\mathcal{E}_{\text{topo}}d\right)=\tilde{\mu}\left(\frac{p}{2}+\frac{q}{6}\right),\label{eq:M}
\end{equation}
where $\tilde{\mu}=\gamma/B$ is the rescaled chemical potential which
can be regarded as a constant for fixed $\alpha$ and $\beta$. Such
magnetization of axial domain walls has been well understood for
e.g., separate $\pi^{0}$ and $\eta$ solitons in a magnetic field,
from the anomalous action coupling $A_{\mu}^{\text{B}}$ and $A_{\mu}$.
Here we generalize this idea to our mixed soliton lattices allowed
by $U$. The nuance is, different $\left(p,q\right)$ feature different
winding numbers and occupy different spatial lengths. Hence it is more
appropriate to consider the average of Eq.~\eqref{eq:M} over the lattice period $d_{L}$, defined by
$\bar{\mathrm{m}}\equiv\mathrm{M}/d_{L}$. 
If we regard $\bar{\mathrm{m}}$ as a function of $\gamma$, it would be a piecewise function with
each jump representing a certain alternation of ground states with changed $p$ and $q$. 
Again we take the five configurations appearing
in Eq.~\eqref{eq:windows} as examples, demonstrating the behavior
of $\bar{\mathrm{m}}/\tilde{\mu}$ in Fig.~\ref{fig:staircase}. 
Such a figure manifests the first order nature of the phase transition between varied ground states.
The $\bar{\mathrm{m}}(\gamma)$ diagram could exhibit finer structures when more possibilities of $(p,q)$ are considered.
We would further delve into this point in future works as briefed in Sec.~\ref{sec:summary}.  

\section{Quasicrystals in Strong Magnetic Field 
and/or High Density Limit}
\label{sec:strong-B}

Hinted by the ground state alternation presented in the previous section,
a nontrivial phenomenon is, despite the weight factor $1/3$ of $q$
in comparison to $p$ among the contribution to $\mathcal{E}_{\text{topo}}\propto-\gamma$,
the ground state for $\gamma\rightarrow\infty$ is not $\left(p,q\right)=\left(2,0\right)$.
In conclusion, we discover the ground state in the strong magnetic
field and/or high density (both mean large $\gamma$) limit is the
mixed soliton configuration satisfying 
\begin{equation}
\text{Ground State at }\gamma\rightarrow\infty\text{:}\quad r\equiv\frac{p}{q}\simeq\frac{3}{\alpha}\label{eq:ratio}
\end{equation}
while being independent of $\beta$. We manifest this fact by comparing
soliton lattice configurations of different ratios $r\equiv p/q$
and finding the $r$ with the lowest $\mathcal{E}$ under a very large
input of $\gamma$. To be specific, we make our enumeration of $p$
and $q$ limited (and therefore practical) by setting the precision
goal of $r$ to $\mathcal{O}\left(1\right)$\footnote{The higher precision is achieved by the larger number of $p$ and
$q$. For example, $r=4.2$ and $4.3$ are realized by $\left(p,q\right)=\left(42,10\right)$
and $\left(86,20\right)$ while $r=4.28$ is realized by $\left(p,q\right)=\left(428,100\right)$.}. We would explain our result Eq.~\eqref{eq:ratio} via  a semi-analytical
approximation scheme that disentangles the mixed soliton EOMs.

\begin{figure}     
\centering       
\includegraphics[width=0.75\columnwidth]{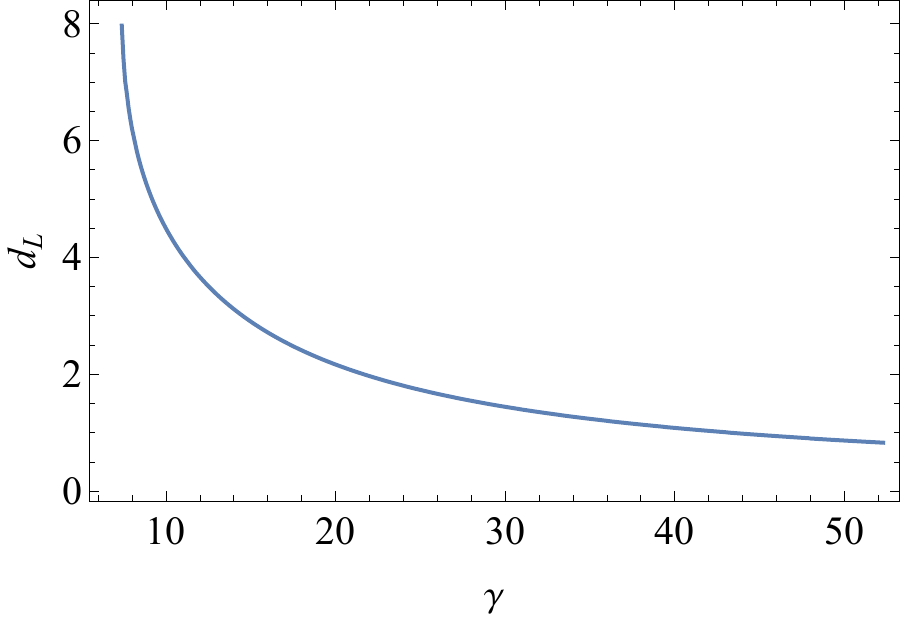}      
\caption{Dependence of the lattice period of configuration $(p,q)=(3,1)$ on $\gamma$.}     
\label{fig:d31}  
\end{figure}

The first important observation relying on our numerical results is:
$d_{L}$ tends to be small when $\gamma$ is large. We illustrate
such a tendency with the case of the configuration $\left(3,1\right)$
under again $\alpha=0.7$ and $\beta=\pi/16$ in Fig.~\ref{fig:d31}.
The shape of $d_{L}\left(\gamma\right)$ remains qualitatively similar
for other values of $\alpha,\beta,p,q$. This fact could already explain
the irrelevance of $\beta$ in Eq.~\eqref{eq:ratio}. Looking into
the energy functional $H$ in Eq.~\eqref{eq:H}, one would find the $\beta$
related term is or order $\mathcal{O}\left(1\right)$ while the kinetic
term and the $\gamma$-related term scale as $\mathcal{O}\left(d^{-2}\right)$
and $\mathcal{O}\left(d^{-1}\right)$ respectively. Therefore when
$d$ is squeezed to a very small quantity by the large $\gamma$,
the $\beta$-related contribution to the energy becomes negligible.
Physically, given the $\beta$-related term is induced by the $\pi^{0}$
and $\eta$ masses, its effect is naturally negligible under a large
chemical potential $\mu\gg m$.

In all, we make the approximation of dropping the $\mathcal{O}\left(d^{0}\right)$
term in the Hamiltonian in Eq.~\eqref{eq:H}:
\begin{equation}
H\left(\gamma\rightarrow\infty\right)\simeq\frac{1}{2}\left(\alpha\phi_{3}^{\prime2}+\phi_{0}^{\prime2}\right)-\frac{\gamma}{2\pi}\left(\phi_{3}^{\prime}+\frac{1}{3}\phi_{0}^{\prime}\right)\equiv H_{\infty}.
\end{equation}
Such an approximate $H_{\infty}$ can be minimized in the following
way on a functional level directly
\begin{align}
H_{\infty} & =\frac{1}{2}\left[\left(\sqrt{\alpha}\phi_{3}^{\prime}-\frac{\gamma}{2\sqrt{\alpha}\pi}\right)^{2}+\left(\phi_{0}^{\prime}-\frac{\gamma}{6\pi}\right)^{2}\right]-\frac{\gamma^{2}}{8\pi^{2}}\left(\frac{1}{\alpha}+\frac{1}{9}\right)\nonumber \\
 & \geq-\frac{\gamma^{2}}{8\pi^{2}}\left(\frac{1}{\alpha}+\frac{1}{9}\right)\equiv E_{\min}.
\end{align}
The minimization is realized when the $\phi_{0,3}$ are solved by
\begin{equation}
\phi_{3}=\frac{\gamma}{2\alpha\pi}z,\quad\phi_{0}=\frac{\gamma}{6\pi}z.
\label{eq:sol}
\end{equation}
One can further derive the lattice period $d_{L}$ by matching the
boundary conditions of $\phi_{0,3}$:
\begin{equation}
d_{L}=\frac{2\pi^{2}}{\gamma}\cdot p\alpha=\frac{2\pi^{2}}{\gamma}\cdot3q 
\quad \Rightarrow \quad 
\frac{p}{q}=\frac{3}{\alpha},
\label{eq:ratio2}
\end{equation}
which in the meantime manifests the ground state ratio stated in Eq.~\eqref{eq:ratio}.
Such a semi-analytic proof yields an interesting interpretation that
in the strong $\gamma$ limit, the mixed soliton lattice, 
is actually decoupled to two separate CSLs of $\phi_{0,3}$ whose EOMs are solved independently.
This is because the nontrivial coupling between them comes from the $\beta$-related
terms that prove irrelevant for large $\gamma$. 
What binds them together is the common lattice period $d$, i.e., the mixed soliton
configuration stipulated by the underlying torus topology $T^{2}=S^{1}\times S^{1}$.
Then naturally, the mixture proportion of two CSLs signified by $r$ depends on the proportion of their energy contribution in $H_{\infty}$, which is ultimately determined by the
proportion of coefficients $\alpha=f_{\pi}^{2}/f_{\eta}^{2}$ in
quadratic terms and the $1/3$ in $\mathcal{L}_{B}$.

\begin{figure}
\centering
\includegraphics[width=0.75\columnwidth]{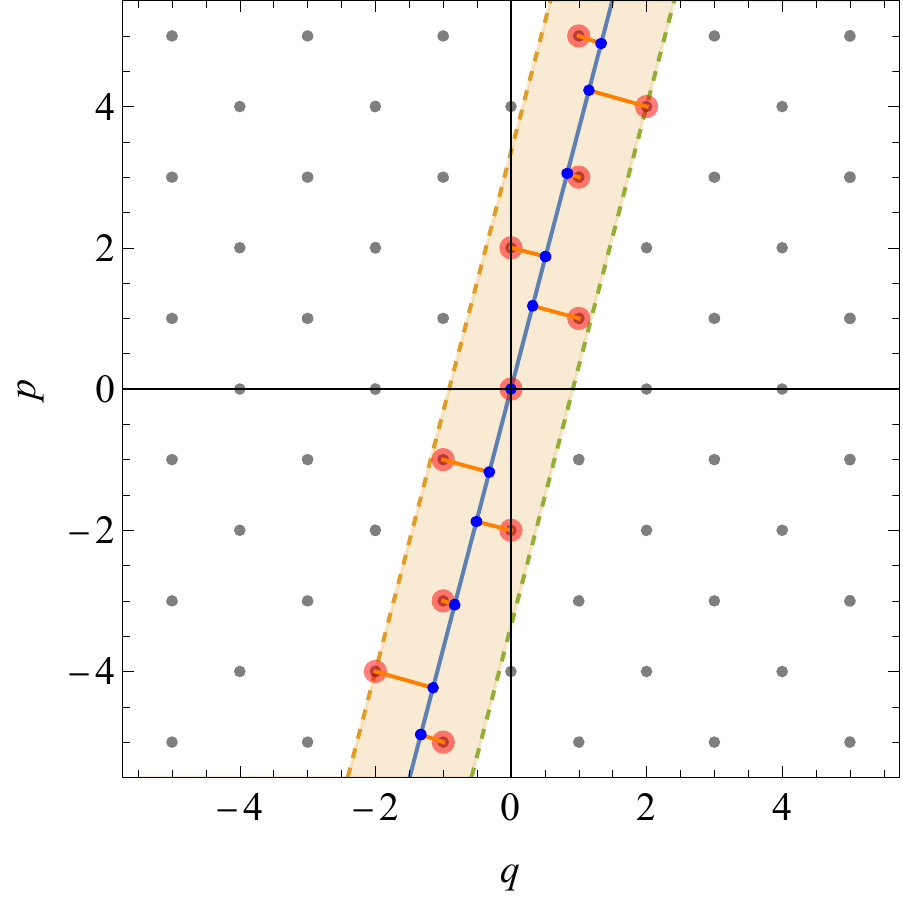}
\caption{Ground state quasicrystal demonstrated on the $p$-$q$ plane. Gray dots represent legitimate mixed soliton lattice satisfying Eq.~\eqref{eq:singlevalue}. The blue line has a slope $3/\alpha$ with $\alpha=0.3e$. The light orange band bounded by orange and green dash lines indicates nearby $(p,q)$ dots. These dots are highlighted with burnt orange color. The solid orange lines mark the distances between 
the $(p,q)$ dots and the $p=3q/\alpha$ slope. The closer such distance is, the more approximate the $(p,q)$ configuration is to the ground state.}
\label{fig:quasi}  
\end{figure} 

The subtlety of Eq.~\eqref{eq:ratio2} lies in that $p/q$ is obviously a rational number
while the  $3/\alpha$ on the right hand side of the equation is generally irrational. 
That means, subject to the condition Eq.~\eqref{eq:singlevalue}, the ground state
could only select the optimal $p$ and $q$ that produces the ratio nearest to 
(irrational) $3/\alpha$. 
The result is thereby a 1D quasicrystal which can be demonstrated vividly 
via the projection onto a 2D grid spanned by 
the $p,q$-axes. 
We call such a quasicrystal the ``chiral soliton quasicrystal'' and show an exemplary case in Fig.~\ref{fig:quasi} with
the choice $\alpha=0.3e$. Each gray dot represents a certain type of mixed soliton lattice. 
The nearest dot (apart from the origin) to the blue line $p=10q/e$ is actually at the infinity of the plane. 
As one summons larger $p$ and $q$,
the rational number $p/q$ approximates 
the irrational number $3/\alpha$ more accurately.  
Or in other words, the periodicity with finite $p$ and $q$ could be realized only in an approximate sense, 
while the true ground state does not have a periodicity.
That is why such a configuration is referred to as a quasicrystal.
To the best of our knowledge, this discovery is the first incarnation of a quasicrystal in QCD.

In this way, with our innovative mixed solitons taken into
account, Eq.~\eqref{eq:ratio2} informs us, at least in strong magnetic field limit, the ground state is a mixed soliton lattice (for rational $\alpha$) or quasicrystal (for irrational $\alpha$), whose critical
$\gamma$ could be lower than that of the $\pi^{0}$-CSL, letting alone the $\eta$-CSL. 
The set of irrational numbers is dense and 
while the set of rational numbers is not dense.
Thus, the ground state is generally 
a chiral soliton quasicrystal.
Such a conclusion also applies to the case with a magnetic field of intermediate strength, depending
on $\alpha$ and $\beta$, as substantiated in Sec.~\ref{sec:GS}. 
These lead to a revolutionary idea that the density for $\eta$ to become physically relevant could be lower than thought, as long as its mixture with $\pi^{0}$ in our scenario is accounted.

Finally, we shall remark on the applicability of our scenario
with respect to the magnitude of $\gamma$. 
Our theoretical framework is based on the perturbative expansion in terms of a small momentum (spatial derivative) compared to the scale of the decay constants.
Specifically, such a scale is the cutoff momentum  $\Lambda \simeq4\pi f_{\pi,\eta}$ brought in by considering loop effects beginning with the quartic term of momentum.  
On the other hand, one can easily read off the characteristic momentum scale of our mixed soliton lattice/quasicrystal in strong $\gamma$ limit from Eq.~\eqref{eq:sol}.
We should guarantee the characteristic momenta to be much smaller than the cutoff momentum so that the derivative terms of higher order in the Lagrangian are irrelevant compared to $\mathcal{L}_B$, maintaining the perturbation hierarchy.
In our dimensionless definitions, especially with the $f_{\pi,\eta}$ absorbed into $\pi_3$ and $\eta$ fields (c.f. Eq.~\eqref{eq:phi}), the constraint on $\gamma$ reads
\begin{equation} \max\left[\frac{\gamma}{2\alpha\pi},\frac{\gamma}{6\pi}\right]\ll4\pi. \end{equation} 
One can easily check that, for instance, the special values of $\gamma$ in Eq.~\eqref{eq:windows} satisfy $\gamma\ll5.6\pi^{2}$ indeed.
Such a constraint could be translated with physical units, $\mu B \ll 16\pi^3 \min [f_\pi^3,f_\eta/3]$, which indicates a window of density and/or magnetic field for the validity of our study.

\section{Conclusion and Outlook} \label{sec:summary} 

In this work, we have introduced the idea of the mixed soliton lattices and quasicrystals 
in the context
of magnetized dense hadronic matter. It has been known that the $\pi^{0}$
and $\eta$ mesons could emerge as domain walls in their corresponding
density regimes. Nonetheless, unlike the previous discussions on them at separate
energy scales, we discover that the ground state could
be a mixture of the two kinds of solitons with a certain proportion described
by $r=p/q$. Such an $r$ would alter depending on the value of $\mu B$, dubbed ``ground state alternation''. 
The magnetization of such
a mixed soliton lattice is also quantified, yielding a piecewise function
of $\mu B$. 
In the high density and/or strong magnetic field limit,
we have highlighted that the ground state tends to be with 
$r_g=3/\alpha=3f_{\eta}^{2}/f_{\pi}^{2}$. If such $r_g$ happens to be a rational number, then a specific type of a mixed soliton lattice with the corresponding finite $p,q\neq0$ would take up the ground state.
If $r_g$ is irrational, the ground state is a mixed soliton quasicrystal. 
Given the densities of 
(ir)rational numbers, 
the ground state is in general 
a chiral soliton quasicrystal.
In either case, the critical magnetic field could be lower than that of a separate $\pi^0$ or $\eta$ CSL.

We have several outlooks for further exploration. One is to consider
charged pions which are ignored portions among $U(2)$
in our current proposal. Of course, they form excitations with higher
energy due to electromagnetic interactions. They would bring in richer
topological structures from a different homotopy class. 
See Ref.~\cite{Eto:2023lyo} for domain-wall Skyrmions.
Another point
we would further delve into is a more quantitative explanation of
the underlying mechanism that reduces the energy of mixed soliton
compared to its individual ingredients. That relies on the analysis of the interaction between $\pi^{0}$ and $\eta$.
Relevant discussions in Ref.~\cite{Eto:2021gyy} can be followed.

Moreover, we have gripping results under discussion for forthcoming works. 
The winding number in our scenario reads $\left(p+q\right)/2$. Our scenario of the ground state alternation implies the ground state winding number as a function of the magnetic field could be a Devil's Staircase (Cantor function), which is novel and peculiar in the context of the QCD ground state. We would delve into this point with concrete evaluations of mass parameters $a$ and $bm$, chasing a realistic context for experimental observations.   

Also significant is the extension study of our quasicrystal, especially the gapless modes.
In the case of usual lattices, the translational symmetry is spontaneously broken to a discrete symmetry, and consequently 
there is a Nambu-Goldstone mode corresponding to the spontaneously broken translational symmetry, a phonon.
On the other hand, 
the translational symmetry is 
completely broken in quasicrystals.
Quasicrystals are known to allow 
diffusive Nambu-Goldstone modes, called phasons 
\cite{Baggioli:2020haa}.
In order to study further aspects of quasicrystals in field theory, 
one should refer 
Refs.~\cite{Baggioli:2020haa,Surowka:2021ved}.
In our case, in addition to them, there should be also ${\mathbb C}P^1$ modes corresponding to the spontaneously 
broken the $SU(2)_{\rm V}$ symmetry down to $U(1)$, as the case of the non-Abelian CSL in a rotation \cite{Eto:2021gyy}.
Whether such non-Abelian modes are also diffusive is an interesting problem to explore.

\begin{acknowledgments}
We thank Thomas Brauner for useful comments.
This work is supported in part by 
 JSPS KAKENHI [Grants No. JP22H01221], and the WPI program ``Sustainability with Knotted Chiral Meta Matter (SKCM$^2$)'' at Hiroshima University.
 Z.Q was supported by JSPS KAKENHI, Grant-in-Aid for Scientific Research No. JP20J20974 (PD).
\end{acknowledgments}

\bibliographystyle{apsrev4-1}

\begin{thebibliography}{50}%
\makeatletter
\providecommand \@ifxundefined [1]{%
 \@ifx{#1\undefined}
}%
\providecommand \@ifnum [1]{%
 \ifnum #1\expandafter \@firstoftwo
 \else \expandafter \@secondoftwo
 \fi
}%
\providecommand \@ifx [1]{%
 \ifx #1\expandafter \@firstoftwo
 \else \expandafter \@secondoftwo
 \fi
}%
\providecommand \natexlab [1]{#1}%
\providecommand \enquote  [1]{``#1''}%
\providecommand \bibnamefont  [1]{#1}%
\providecommand \bibfnamefont [1]{#1}%
\providecommand \citenamefont [1]{#1}%
\providecommand \href@noop [0]{\@secondoftwo}%
\providecommand \href [0]{\begingroup \@sanitize@url \@href}%
\providecommand \@href[1]{\@@startlink{#1}\@@href}%
\providecommand \@@href[1]{\endgroup#1\@@endlink}%
\providecommand \@sanitize@url [0]{\catcode `\\12\catcode `\$12\catcode
  `\&12\catcode `\#12\catcode `\^12\catcode `\_12\catcode `\%12\relax}%
\providecommand \@@startlink[1]{}%
\providecommand \@@endlink[0]{}%
\providecommand \url  [0]{\begingroup\@sanitize@url \@url }%
\providecommand \@url [1]{\endgroup\@href {#1}{\urlprefix }}%
\providecommand \urlprefix  [0]{URL }%
\providecommand \Eprint [0]{\href }%
\providecommand \doibase [0]{http://dx.doi.org/}%
\providecommand \selectlanguage [0]{\@gobble}%
\providecommand \bibinfo  [0]{\@secondoftwo}%
\providecommand \bibfield  [0]{\@secondoftwo}%
\providecommand \translation [1]{[#1]}%
\providecommand \BibitemOpen [0]{}%
\providecommand \bibitemStop [0]{}%
\providecommand \bibitemNoStop [0]{.\EOS\space}%
\providecommand \EOS [0]{\spacefactor3000\relax}%
\providecommand \BibitemShut  [1]{\csname bibitem#1\endcsname}%
\let\auto@bib@innerbib\@empty
\bibitem [{\citenamefont {Son}\ and\ \citenamefont
  {Stephanov}(2008)}]{Son:2007ny}%
  \BibitemOpen
  \bibfield  {author} {\bibinfo {author} {\bibfnamefont {D.~T.}\ \bibnamefont
  {Son}}\ and\ \bibinfo {author} {\bibfnamefont {M.~A.}\ \bibnamefont
  {Stephanov}},\ }\href {\doibase 10.1103/PhysRevD.77.014021} {\bibfield
  {journal} {\bibinfo  {journal} {Phys. Rev. D}\ }\textbf {\bibinfo {volume}
  {77}},\ \bibinfo {pages} {014021} (\bibinfo {year} {2008})},\ \Eprint
  {http://arxiv.org/abs/0710.1084} {arXiv:0710.1084 [hep-ph]} \BibitemShut
  {NoStop}%
\bibitem [{\citenamefont {Eto}\ \emph {et~al.}(2013)\citenamefont {Eto},
  \citenamefont {Hashimoto},\ and\ \citenamefont {Hatsuda}}]{Eto:2012qd}%
  \BibitemOpen
  \bibfield  {author} {\bibinfo {author} {\bibfnamefont {M.}~\bibnamefont
  {Eto}}, \bibinfo {author} {\bibfnamefont {K.}~\bibnamefont {Hashimoto}}, \
  and\ \bibinfo {author} {\bibfnamefont {T.}~\bibnamefont {Hatsuda}},\ }\href
  {\doibase 10.1103/PhysRevD.88.081701} {\bibfield  {journal} {\bibinfo
  {journal} {Phys. Rev. D}\ }\textbf {\bibinfo {volume} {88}},\ \bibinfo
  {pages} {081701} (\bibinfo {year} {2013})},\ \Eprint
  {http://arxiv.org/abs/1209.4814} {arXiv:1209.4814 [hep-ph]} \BibitemShut
  {NoStop}%
\bibitem [{\citenamefont {Brauner}\ and\ \citenamefont
  {Yamamoto}(2017)}]{Brauner:2016pko}%
  \BibitemOpen
  \bibfield  {author} {\bibinfo {author} {\bibfnamefont {T.}~\bibnamefont
  {Brauner}}\ and\ \bibinfo {author} {\bibfnamefont {N.}~\bibnamefont
  {Yamamoto}},\ }\href {\doibase 10.1007/JHEP04(2017)132} {\bibfield  {journal}
  {\bibinfo  {journal} {JHEP}\ }\textbf {\bibinfo {volume} {04}},\ \bibinfo
  {pages} {132} (\bibinfo {year} {2017})},\ \Eprint
  {http://arxiv.org/abs/1609.05213} {arXiv:1609.05213 [hep-ph]} \BibitemShut
  {NoStop}%
\bibitem [{\citenamefont {Brauner}\ \emph
  {et~al.}(2019{\natexlab{a}})\citenamefont {Brauner}, \citenamefont {Filios},\
  and\ \citenamefont {Kole\v{s}ov\'a}}]{Brauner:2019aid}%
  \BibitemOpen
  \bibfield  {author} {\bibinfo {author} {\bibfnamefont {T.}~\bibnamefont
  {Brauner}}, \bibinfo {author} {\bibfnamefont {G.}~\bibnamefont {Filios}}, \
  and\ \bibinfo {author} {\bibfnamefont {H.}~\bibnamefont {Kole\v{s}ov\'a}},\
  }\href {\doibase 10.1007/JHEP12(2019)029} {\bibfield  {journal} {\bibinfo
  {journal} {JHEP}\ }\textbf {\bibinfo {volume} {12}},\ \bibinfo {pages} {029}
  (\bibinfo {year} {2019}{\natexlab{a}})},\ \Eprint
  {http://arxiv.org/abs/1905.11409} {arXiv:1905.11409 [hep-ph]} \BibitemShut
  {NoStop}%
\bibitem [{\citenamefont {Brauner}\ \emph {et~al.}(2021)\citenamefont
  {Brauner}, \citenamefont {Kole\v{s}ov\'a},\ and\ \citenamefont
  {Yamamoto}}]{Brauner:2021sci}%
  \BibitemOpen
  \bibfield  {author} {\bibinfo {author} {\bibfnamefont {T.}~\bibnamefont
  {Brauner}}, \bibinfo {author} {\bibfnamefont {H.}~\bibnamefont
  {Kole\v{s}ov\'a}}, \ and\ \bibinfo {author} {\bibfnamefont {N.}~\bibnamefont
  {Yamamoto}},\ }\href {\doibase 10.1016/j.physletb.2021.136767} {\bibfield
  {journal} {\bibinfo  {journal} {Phys. Lett. B}\ }\textbf {\bibinfo {volume}
  {823}},\ \bibinfo {pages} {136767} (\bibinfo {year} {2021})},\ \Eprint
  {http://arxiv.org/abs/2108.10044} {arXiv:2108.10044 [hep-ph]} \BibitemShut
  {NoStop}%
\bibitem [{\citenamefont {Evans}\ and\ \citenamefont
  {Schmitt}(2022)}]{Evans:2022hwr}%
  \BibitemOpen
  \bibfield  {author} {\bibinfo {author} {\bibfnamefont {G.~W.}\ \bibnamefont
  {Evans}}\ and\ \bibinfo {author} {\bibfnamefont {A.}~\bibnamefont
  {Schmitt}},\ }\href {\doibase 10.1007/JHEP09(2022)192} {\bibfield  {journal}
  {\bibinfo  {journal} {JHEP}\ }\textbf {\bibinfo {volume} {09}},\ \bibinfo
  {pages} {192} (\bibinfo {year} {2022})},\ \Eprint
  {http://arxiv.org/abs/2206.01227} {arXiv:2206.01227 [hep-th]} \BibitemShut
  {NoStop}%
\bibitem [{\citenamefont {Gr\o{}nli}\ and\ \citenamefont
  {Brauner}(2022)}]{Gronli:2022cri}%
  \BibitemOpen
  \bibfield  {author} {\bibinfo {author} {\bibfnamefont {M.~S.}\ \bibnamefont
  {Gr\o{}nli}}\ and\ \bibinfo {author} {\bibfnamefont {T.}~\bibnamefont
  {Brauner}},\ }\href {\doibase 10.1140/epjc/s10052-022-10300-5} {\bibfield
  {journal} {\bibinfo  {journal} {Eur. Phys. J. C}\ }\textbf {\bibinfo {volume}
  {82}},\ \bibinfo {pages} {354} (\bibinfo {year} {2022})},\ \Eprint
  {http://arxiv.org/abs/2201.07065} {arXiv:2201.07065 [hep-ph]} \BibitemShut
  {NoStop}%
\bibitem [{\citenamefont {Brauner}\ and\ \citenamefont
  {Kadam}(2017{\natexlab{a}})}]{Brauner:2017uiu}%
  \BibitemOpen
  \bibfield  {author} {\bibinfo {author} {\bibfnamefont {T.}~\bibnamefont
  {Brauner}}\ and\ \bibinfo {author} {\bibfnamefont {S.~V.}\ \bibnamefont
  {Kadam}},\ }\href {\doibase 10.1007/JHEP11(2017)103} {\bibfield  {journal}
  {\bibinfo  {journal} {JHEP}\ }\textbf {\bibinfo {volume} {11}},\ \bibinfo
  {pages} {103} (\bibinfo {year} {2017}{\natexlab{a}})},\ \Eprint
  {http://arxiv.org/abs/1706.04514} {arXiv:1706.04514 [hep-ph]} \BibitemShut
  {NoStop}%
\bibitem [{\citenamefont {Brauner}\ and\ \citenamefont
  {Kadam}(2017{\natexlab{b}})}]{Brauner:2017mui}%
  \BibitemOpen
  \bibfield  {author} {\bibinfo {author} {\bibfnamefont {T.}~\bibnamefont
  {Brauner}}\ and\ \bibinfo {author} {\bibfnamefont {S.}~\bibnamefont
  {Kadam}},\ }\href {\doibase 10.1007/JHEP03(2017)015} {\bibfield  {journal}
  {\bibinfo  {journal} {JHEP}\ }\textbf {\bibinfo {volume} {03}},\ \bibinfo
  {pages} {015} (\bibinfo {year} {2017}{\natexlab{b}})},\ \Eprint
  {http://arxiv.org/abs/1701.06793} {arXiv:1701.06793 [hep-ph]} \BibitemShut
  {NoStop}%
\bibitem [{\citenamefont {Brauner}\ and\ \citenamefont
  {Kole\v{s}ov\'a}(2023)}]{Brauner:2023ort}%
  \BibitemOpen
  \bibfield  {author} {\bibinfo {author} {\bibfnamefont {T.}~\bibnamefont
  {Brauner}}\ and\ \bibinfo {author} {\bibfnamefont {H.}~\bibnamefont
  {Kole\v{s}ov\'a}},\ }\href@noop {} {\  (\bibinfo {year} {2023})},\ \Eprint
  {http://arxiv.org/abs/2302.06902} {arXiv:2302.06902 [hep-ph]} \BibitemShut
  {NoStop}%
\bibitem [{\citenamefont {Huang}\ \emph {et~al.}(2018)\citenamefont {Huang},
  \citenamefont {Nishimura},\ and\ \citenamefont {Yamamoto}}]{Huang:2017pqe}%
  \BibitemOpen
  \bibfield  {author} {\bibinfo {author} {\bibfnamefont {X.-G.}\ \bibnamefont
  {Huang}}, \bibinfo {author} {\bibfnamefont {K.}~\bibnamefont {Nishimura}}, \
  and\ \bibinfo {author} {\bibfnamefont {N.}~\bibnamefont {Yamamoto}},\ }\href
  {\doibase 10.1007/JHEP02(2018)069} {\bibfield  {journal} {\bibinfo  {journal}
  {JHEP}\ }\textbf {\bibinfo {volume} {02}},\ \bibinfo {pages} {069} (\bibinfo
  {year} {2018})},\ \Eprint {http://arxiv.org/abs/1711.02190} {arXiv:1711.02190
  [hep-ph]} \BibitemShut {NoStop}%
\bibitem [{\citenamefont {Nishimura}\ and\ \citenamefont
  {Yamamoto}(2020)}]{Nishimura:2020odq}%
  \BibitemOpen
  \bibfield  {author} {\bibinfo {author} {\bibfnamefont {K.}~\bibnamefont
  {Nishimura}}\ and\ \bibinfo {author} {\bibfnamefont {N.}~\bibnamefont
  {Yamamoto}},\ }\href {\doibase 10.1007/JHEP07(2020)196} {\bibfield  {journal}
  {\bibinfo  {journal} {JHEP}\ }\textbf {\bibinfo {volume} {07}},\ \bibinfo
  {pages} {196} (\bibinfo {year} {2020})},\ \Eprint
  {http://arxiv.org/abs/2003.13945} {arXiv:2003.13945 [hep-ph]} \BibitemShut
  {NoStop}%
\bibitem [{\citenamefont {Eto}\ \emph {et~al.}(2022)\citenamefont {Eto},
  \citenamefont {Nishimura},\ and\ \citenamefont {Nitta}}]{Eto:2021gyy}%
  \BibitemOpen
  \bibfield  {author} {\bibinfo {author} {\bibfnamefont {M.}~\bibnamefont
  {Eto}}, \bibinfo {author} {\bibfnamefont {K.}~\bibnamefont {Nishimura}}, \
  and\ \bibinfo {author} {\bibfnamefont {M.}~\bibnamefont {Nitta}},\ }\href
  {\doibase 10.1007/JHEP08(2022)305} {\bibfield  {journal} {\bibinfo  {journal}
  {JHEP}\ }\textbf {\bibinfo {volume} {08}},\ \bibinfo {pages} {305} (\bibinfo
  {year} {2022})},\ \Eprint {http://arxiv.org/abs/2112.01381} {arXiv:2112.01381
  [hep-ph]} \BibitemShut {NoStop}%
\bibitem [{\citenamefont {Chen}\ \emph {et~al.}(2021)\citenamefont {Chen},
  \citenamefont {Huang},\ and\ \citenamefont {Liao}}]{Chen:2021aiq}%
  \BibitemOpen
  \bibfield  {author} {\bibinfo {author} {\bibfnamefont {H.-L.}\ \bibnamefont
  {Chen}}, \bibinfo {author} {\bibfnamefont {X.-G.}\ \bibnamefont {Huang}}, \
  and\ \bibinfo {author} {\bibfnamefont {J.}~\bibnamefont {Liao}},\ }\href
  {\doibase 10.1007/978-3-030-71427-7_11} {\bibfield  {journal} {\bibinfo
  {journal} {Lect. Notes Phys.}\ }\textbf {\bibinfo {volume} {987}},\ \bibinfo
  {pages} {349} (\bibinfo {year} {2021})},\ \Eprint
  {http://arxiv.org/abs/2108.00586} {arXiv:2108.00586 [hep-ph]} \BibitemShut
  {NoStop}%
\bibitem [{\citenamefont {Eto}\ and\ \citenamefont
  {Nitta}(2022)}]{Eto:2022lhu}%
  \BibitemOpen
  \bibfield  {author} {\bibinfo {author} {\bibfnamefont {M.}~\bibnamefont
  {Eto}}\ and\ \bibinfo {author} {\bibfnamefont {M.}~\bibnamefont {Nitta}},\
  }\href {\doibase 10.1007/JHEP09(2022)077} {\bibfield  {journal} {\bibinfo
  {journal} {JHEP}\ }\textbf {\bibinfo {volume} {09}},\ \bibinfo {pages} {077}
  (\bibinfo {year} {2022})},\ \Eprint {http://arxiv.org/abs/2207.00211}
  {arXiv:2207.00211 [hep-th]} \BibitemShut {NoStop}%
\bibitem [{\citenamefont {Higaki}\ \emph {et~al.}(2022)\citenamefont {Higaki},
  \citenamefont {Kamada},\ and\ \citenamefont {Nishimura}}]{Higaki:2022gnw}%
  \BibitemOpen
  \bibfield  {author} {\bibinfo {author} {\bibfnamefont {T.}~\bibnamefont
  {Higaki}}, \bibinfo {author} {\bibfnamefont {K.}~\bibnamefont {Kamada}}, \
  and\ \bibinfo {author} {\bibfnamefont {K.}~\bibnamefont {Nishimura}},\ }\href
  {\doibase 10.1103/PhysRevD.106.096022} {\bibfield  {journal} {\bibinfo
  {journal} {Phys. Rev. D}\ }\textbf {\bibinfo {volume} {106}},\ \bibinfo
  {pages} {096022} (\bibinfo {year} {2022})},\ \Eprint
  {http://arxiv.org/abs/2207.00212} {arXiv:2207.00212 [hep-th]} \BibitemShut
  {NoStop}%
\bibitem [{\citenamefont {Yamada}\ and\ \citenamefont
  {Yamamoto}(2021)}]{Yamada:2021jhy}%
  \BibitemOpen
  \bibfield  {author} {\bibinfo {author} {\bibfnamefont {A.}~\bibnamefont
  {Yamada}}\ and\ \bibinfo {author} {\bibfnamefont {N.}~\bibnamefont
  {Yamamoto}},\ }\href {\doibase 10.1103/PhysRevD.104.054041} {\bibfield
  {journal} {\bibinfo  {journal} {Phys. Rev. D}\ }\textbf {\bibinfo {volume}
  {104}},\ \bibinfo {pages} {054041} (\bibinfo {year} {2021})},\ \Eprint
  {http://arxiv.org/abs/2107.07074} {arXiv:2107.07074 [hep-ph]} \BibitemShut
  {NoStop}%
\bibitem [{\citenamefont {Brauner}\ \emph
  {et~al.}(2019{\natexlab{b}})\citenamefont {Brauner}, \citenamefont {Filios},\
  and\ \citenamefont {Kole\v{s}ov\'a}}]{Brauner:2019rjg}%
  \BibitemOpen
  \bibfield  {author} {\bibinfo {author} {\bibfnamefont {T.}~\bibnamefont
  {Brauner}}, \bibinfo {author} {\bibfnamefont {G.}~\bibnamefont {Filios}}, \
  and\ \bibinfo {author} {\bibfnamefont {H.}~\bibnamefont {Kole\v{s}ov\'a}},\
  }\href {\doibase 10.1103/PhysRevLett.123.012001} {\bibfield  {journal}
  {\bibinfo  {journal} {Phys. Rev. Lett.}\ }\textbf {\bibinfo {volume} {123}},\
  \bibinfo {pages} {012001} (\bibinfo {year} {2019}{\natexlab{b}})},\ \Eprint
  {http://arxiv.org/abs/1902.07522} {arXiv:1902.07522 [hep-ph]} \BibitemShut
  {NoStop}%
\bibitem [{\citenamefont {Son}\ and\ \citenamefont
  {Zhitnitsky}(2004)}]{Son:2004tq}%
  \BibitemOpen
  \bibfield  {author} {\bibinfo {author} {\bibfnamefont {D.~T.}\ \bibnamefont
  {Son}}\ and\ \bibinfo {author} {\bibfnamefont {A.~R.}\ \bibnamefont
  {Zhitnitsky}},\ }\href {\doibase 10.1103/PhysRevD.70.074018} {\bibfield
  {journal} {\bibinfo  {journal} {Phys. Rev. D}\ }\textbf {\bibinfo {volume}
  {70}},\ \bibinfo {pages} {074018} (\bibinfo {year} {2004})},\ \Eprint
  {http://arxiv.org/abs/hep-ph/0405216} {arXiv:hep-ph/0405216} \BibitemShut
  {NoStop}%
\bibitem [{\citenamefont {Chen}\ \emph {et~al.}(2022)\citenamefont {Chen},
  \citenamefont {Fukushima},\ and\ \citenamefont {Qiu}}]{Chen:2021vou}%
  \BibitemOpen
  \bibfield  {author} {\bibinfo {author} {\bibfnamefont {S.}~\bibnamefont
  {Chen}}, \bibinfo {author} {\bibfnamefont {K.}~\bibnamefont {Fukushima}}, \
  and\ \bibinfo {author} {\bibfnamefont {Z.}~\bibnamefont {Qiu}},\ }\href
  {\doibase 10.1103/PhysRevD.105.L011502} {\bibfield  {journal} {\bibinfo
  {journal} {Phys. Rev. D}\ }\textbf {\bibinfo {volume} {105}},\ \bibinfo
  {pages} {L011502} (\bibinfo {year} {2022})},\ \Eprint
  {http://arxiv.org/abs/2104.11482} {arXiv:2104.11482 [hep-ph]} \BibitemShut
  {NoStop}%
\bibitem [{\citenamefont {Chen}\ \emph {et~al.}(2023)\citenamefont {Chen},
  \citenamefont {Fukushima},\ and\ \citenamefont {Qiu}}]{Chen:2023jbq}%
  \BibitemOpen
  \bibfield  {author} {\bibinfo {author} {\bibfnamefont {S.}~\bibnamefont
  {Chen}}, \bibinfo {author} {\bibfnamefont {K.}~\bibnamefont {Fukushima}}, \
  and\ \bibinfo {author} {\bibfnamefont {Z.}~\bibnamefont {Qiu}},\ }\href@noop
  {} {\  (\bibinfo {year} {2023})},\ \Eprint {http://arxiv.org/abs/2303.04692}
  {arXiv:2303.04692 [hep-th]} \BibitemShut {NoStop}%
\bibitem [{\citenamefont {Kawaguchi}\ \emph {et~al.}(2019)\citenamefont
  {Kawaguchi}, \citenamefont {Ma},\ and\ \citenamefont
  {Matsuzaki}}]{Kawaguchi:2018fpi}%
  \BibitemOpen
  \bibfield  {author} {\bibinfo {author} {\bibfnamefont {M.}~\bibnamefont
  {Kawaguchi}}, \bibinfo {author} {\bibfnamefont {Y.-L.}\ \bibnamefont {Ma}}, \
  and\ \bibinfo {author} {\bibfnamefont {S.}~\bibnamefont {Matsuzaki}},\ }\href
  {\doibase 10.1103/PhysRevC.100.025207} {\bibfield  {journal} {\bibinfo
  {journal} {Phys. Rev. C}\ }\textbf {\bibinfo {volume} {100}},\ \bibinfo
  {pages} {025207} (\bibinfo {year} {2019})},\ \Eprint
  {http://arxiv.org/abs/1810.12880} {arXiv:1810.12880 [nucl-th]} \BibitemShut
  {NoStop}%
\bibitem [{\citenamefont {Eto}\ \emph {et~al.}(2023)\citenamefont {Eto},
  \citenamefont {Nishimura},\ and\ \citenamefont {Nitta}}]{Eto:2023lyo}%
  \BibitemOpen
  \bibfield  {author} {\bibinfo {author} {\bibfnamefont {M.}~\bibnamefont
  {Eto}}, \bibinfo {author} {\bibfnamefont {K.}~\bibnamefont {Nishimura}}, \
  and\ \bibinfo {author} {\bibfnamefont {M.}~\bibnamefont {Nitta}},\
  }\href@noop {} {\  (\bibinfo {year} {2023})},\ \Eprint
  {http://arxiv.org/abs/2304.02940} {arXiv:2304.02940 [hep-ph]} \BibitemShut
  {NoStop}%
\bibitem [{\citenamefont {Nitta}(2013)}]{Nitta:2012wi}%
  \BibitemOpen
  \bibfield  {author} {\bibinfo {author} {\bibfnamefont {M.}~\bibnamefont
  {Nitta}},\ }\href {\doibase 10.1103/PhysRevD.87.025013} {\bibfield  {journal}
  {\bibinfo  {journal} {Phys. Rev. D}\ }\textbf {\bibinfo {volume} {87}},\
  \bibinfo {pages} {025013} (\bibinfo {year} {2013})},\ \Eprint
  {http://arxiv.org/abs/1210.2233} {arXiv:1210.2233 [hep-th]} \BibitemShut
  {NoStop}%
\bibitem [{\citenamefont {Nitta}(2022)}]{Nitta:2022ahj}%
  \BibitemOpen
  \bibfield  {author} {\bibinfo {author} {\bibfnamefont {M.}~\bibnamefont
  {Nitta}},\ }\href {\doibase 10.1103/PhysRevD.105.105006} {\bibfield
  {journal} {\bibinfo  {journal} {Phys. Rev. D}\ }\textbf {\bibinfo {volume}
  {105}},\ \bibinfo {pages} {105006} (\bibinfo {year} {2022})},\ \Eprint
  {http://arxiv.org/abs/2202.03929} {arXiv:2202.03929 [hep-th]} \BibitemShut
  {NoStop}%
\bibitem [{\citenamefont {Ross}\ and\ \citenamefont
  {Nitta}(2023)}]{Ross:2022vsa}%
  \BibitemOpen
  \bibfield  {author} {\bibinfo {author} {\bibfnamefont {C.}~\bibnamefont
  {Ross}}\ and\ \bibinfo {author} {\bibfnamefont {M.}~\bibnamefont {Nitta}},\
  }\href {\doibase 10.1103/PhysRevB.107.024422} {\bibfield  {journal} {\bibinfo
   {journal} {Phys. Rev. B}\ }\textbf {\bibinfo {volume} {107}},\ \bibinfo
  {pages} {024422} (\bibinfo {year} {2023})},\ \Eprint
  {http://arxiv.org/abs/2205.11417} {arXiv:2205.11417 [cond-mat.mes-hall]}
  \BibitemShut {NoStop}%
\bibitem [{\citenamefont {Alford}\ \emph {et~al.}(1998)\citenamefont {Alford},
  \citenamefont {Rajagopal},\ and\ \citenamefont {Wilczek}}]{Alford:1997zt}%
  \BibitemOpen
  \bibfield  {author} {\bibinfo {author} {\bibfnamefont {M.~G.}\ \bibnamefont
  {Alford}}, \bibinfo {author} {\bibfnamefont {K.}~\bibnamefont {Rajagopal}}, \
  and\ \bibinfo {author} {\bibfnamefont {F.}~\bibnamefont {Wilczek}},\ }\href
  {\doibase 10.1016/S0370-2693(98)00051-3} {\bibfield  {journal} {\bibinfo
  {journal} {Phys. Lett. B}\ }\textbf {\bibinfo {volume} {422}},\ \bibinfo
  {pages} {247} (\bibinfo {year} {1998})},\ \Eprint
  {http://arxiv.org/abs/hep-ph/9711395} {arXiv:hep-ph/9711395} \BibitemShut
  {NoStop}%
\bibitem [{\citenamefont {Rapp}\ \emph {et~al.}(1998)\citenamefont {Rapp},
  \citenamefont {Sch\"afer}, \citenamefont {Shuryak},\ and\ \citenamefont
  {Velkovsky}}]{Rapp:1997zu}%
  \BibitemOpen
  \bibfield  {author} {\bibinfo {author} {\bibfnamefont {R.}~\bibnamefont
  {Rapp}}, \bibinfo {author} {\bibfnamefont {T.}~\bibnamefont {Sch\"afer}},
  \bibinfo {author} {\bibfnamefont {E.~V.}\ \bibnamefont {Shuryak}}, \ and\
  \bibinfo {author} {\bibfnamefont {M.}~\bibnamefont {Velkovsky}},\ }\href
  {\doibase 10.1103/PhysRevLett.81.53} {\bibfield  {journal} {\bibinfo
  {journal} {Phys. Rev. Lett.}\ }\textbf {\bibinfo {volume} {81}},\ \bibinfo
  {pages} {53} (\bibinfo {year} {1998})},\ \Eprint
  {http://arxiv.org/abs/hep-ph/9711396} {arXiv:hep-ph/9711396} \BibitemShut
  {NoStop}%
\bibitem [{\citenamefont {Alford}\ \emph {et~al.}(1999)\citenamefont {Alford},
  \citenamefont {Rajagopal},\ and\ \citenamefont {Wilczek}}]{Alford:1998mk}%
  \BibitemOpen
  \bibfield  {author} {\bibinfo {author} {\bibfnamefont {M.~G.}\ \bibnamefont
  {Alford}}, \bibinfo {author} {\bibfnamefont {K.}~\bibnamefont {Rajagopal}}, \
  and\ \bibinfo {author} {\bibfnamefont {F.}~\bibnamefont {Wilczek}},\ }\href
  {\doibase 10.1016/S0550-3213(98)00668-3} {\bibfield  {journal} {\bibinfo
  {journal} {Nucl. Phys. B}\ }\textbf {\bibinfo {volume} {537}},\ \bibinfo
  {pages} {443} (\bibinfo {year} {1999})},\ \Eprint
  {http://arxiv.org/abs/hep-ph/9804403} {arXiv:hep-ph/9804403} \BibitemShut
  {NoStop}%
\bibitem [{\citenamefont {Aoki}\ and\ \citenamefont
  {Creutz}(2014)}]{Aoki:2014moa}%
  \BibitemOpen
  \bibfield  {author} {\bibinfo {author} {\bibfnamefont {S.}~\bibnamefont
  {Aoki}}\ and\ \bibinfo {author} {\bibfnamefont {M.}~\bibnamefont {Creutz}},\
  }\href {\doibase 10.1103/PhysRevLett.112.141603} {\bibfield  {journal}
  {\bibinfo  {journal} {Phys. Rev. Lett.}\ }\textbf {\bibinfo {volume} {112}},\
  \bibinfo {pages} {141603} (\bibinfo {year} {2014})},\ \Eprint
  {http://arxiv.org/abs/1402.1837} {arXiv:1402.1837 [hep-lat]} \BibitemShut
  {NoStop}%
\bibitem [{\citenamefont {Nitta}(2015)}]{Nitta:2014rxa}%
  \BibitemOpen
  \bibfield  {author} {\bibinfo {author} {\bibfnamefont {M.}~\bibnamefont
  {Nitta}},\ }\href {\doibase 10.1016/j.nuclphysb.2015.04.006} {\bibfield
  {journal} {\bibinfo  {journal} {Nucl. Phys. B}\ }\textbf {\bibinfo {volume}
  {895}},\ \bibinfo {pages} {288} (\bibinfo {year} {2015})},\ \Eprint
  {http://arxiv.org/abs/1412.8276} {arXiv:1412.8276 [hep-th]} \BibitemShut
  {NoStop}%
\bibitem [{\citenamefont {Eto}\ and\ \citenamefont
  {Nitta}(2015)}]{Eto:2015uqa}%
  \BibitemOpen
  \bibfield  {author} {\bibinfo {author} {\bibfnamefont {M.}~\bibnamefont
  {Eto}}\ and\ \bibinfo {author} {\bibfnamefont {M.}~\bibnamefont {Nitta}},\
  }\href {\doibase 10.1103/PhysRevD.91.085044} {\bibfield  {journal} {\bibinfo
  {journal} {Phys. Rev. D}\ }\textbf {\bibinfo {volume} {91}},\ \bibinfo
  {pages} {085044} (\bibinfo {year} {2015})},\ \Eprint
  {http://arxiv.org/abs/1501.07038} {arXiv:1501.07038 [hep-th]} \BibitemShut
  {NoStop}%
\bibitem [{\citenamefont {Nakano}\ and\ \citenamefont
  {Tatsumi}(2005)}]{Nakano:2004cd}%
  \BibitemOpen
  \bibfield  {author} {\bibinfo {author} {\bibfnamefont {E.}~\bibnamefont
  {Nakano}}\ and\ \bibinfo {author} {\bibfnamefont {T.}~\bibnamefont
  {Tatsumi}},\ }\href {\doibase 10.1103/PhysRevD.71.114006} {\bibfield
  {journal} {\bibinfo  {journal} {Phys. Rev. D}\ }\textbf {\bibinfo {volume}
  {71}},\ \bibinfo {pages} {114006} (\bibinfo {year} {2005})},\ \Eprint
  {http://arxiv.org/abs/hep-ph/0411350} {arXiv:hep-ph/0411350} \BibitemShut
  {NoStop}%
\bibitem [{\citenamefont {Nickel}(2009)}]{Nickel:2009ke}%
  \BibitemOpen
  \bibfield  {author} {\bibinfo {author} {\bibfnamefont {D.}~\bibnamefont
  {Nickel}},\ }\href {\doibase 10.1103/PhysRevLett.103.072301} {\bibfield
  {journal} {\bibinfo  {journal} {Phys. Rev. Lett.}\ }\textbf {\bibinfo
  {volume} {103}},\ \bibinfo {pages} {072301} (\bibinfo {year} {2009})},\
  \Eprint {http://arxiv.org/abs/0902.1778} {arXiv:0902.1778 [hep-ph]}
  \BibitemShut {NoStop}%
\bibitem [{\citenamefont {Basar}\ \emph {et~al.}(2009)\citenamefont {Basar},
  \citenamefont {Dunne},\ and\ \citenamefont {Thies}}]{Basar:2009fg}%
  \BibitemOpen
  \bibfield  {author} {\bibinfo {author} {\bibfnamefont {G.}~\bibnamefont
  {Basar}}, \bibinfo {author} {\bibfnamefont {G.~V.}\ \bibnamefont {Dunne}}, \
  and\ \bibinfo {author} {\bibfnamefont {M.}~\bibnamefont {Thies}},\ }\href
  {\doibase 10.1103/PhysRevD.79.105012} {\bibfield  {journal} {\bibinfo
  {journal} {Phys. Rev. D}\ }\textbf {\bibinfo {volume} {79}},\ \bibinfo
  {pages} {105012} (\bibinfo {year} {2009})},\ \Eprint
  {http://arxiv.org/abs/0903.1868} {arXiv:0903.1868 [hep-th]} \BibitemShut
  {NoStop}%
\bibitem [{\citenamefont {Buballa}\ and\ \citenamefont
  {Carignano}(2015)}]{Buballa:2014tba}%
  \BibitemOpen
  \bibfield  {author} {\bibinfo {author} {\bibfnamefont {M.}~\bibnamefont
  {Buballa}}\ and\ \bibinfo {author} {\bibfnamefont {S.}~\bibnamefont
  {Carignano}},\ }\href {\doibase 10.1016/j.ppnp.2014.11.001} {\bibfield
  {journal} {\bibinfo  {journal} {Prog. Part. Nucl. Phys.}\ }\textbf {\bibinfo
  {volume} {81}},\ \bibinfo {pages} {39} (\bibinfo {year} {2015})},\ \Eprint
  {http://arxiv.org/abs/1406.1367} {arXiv:1406.1367 [hep-ph]} \BibitemShut
  {NoStop}%
\bibitem [{\citenamefont {Hidaka}\ \emph {et~al.}(2015)\citenamefont {Hidaka},
  \citenamefont {Kamikado}, \citenamefont {Kanazawa},\ and\ \citenamefont
  {Noumi}}]{Hidaka:2015xza}%
  \BibitemOpen
  \bibfield  {author} {\bibinfo {author} {\bibfnamefont {Y.}~\bibnamefont
  {Hidaka}}, \bibinfo {author} {\bibfnamefont {K.}~\bibnamefont {Kamikado}},
  \bibinfo {author} {\bibfnamefont {T.}~\bibnamefont {Kanazawa}}, \ and\
  \bibinfo {author} {\bibfnamefont {T.}~\bibnamefont {Noumi}},\ }\href
  {\doibase 10.1103/PhysRevD.92.034003} {\bibfield  {journal} {\bibinfo
  {journal} {Phys. Rev. D}\ }\textbf {\bibinfo {volume} {92}},\ \bibinfo
  {pages} {034003} (\bibinfo {year} {2015})},\ \Eprint
  {http://arxiv.org/abs/1505.00848} {arXiv:1505.00848 [hep-ph]} \BibitemShut
  {NoStop}%
\bibitem [{\citenamefont {Casalbuoni}\ and\ \citenamefont
  {Nardulli}(2004)}]{Casalbuoni:2003wh}%
  \BibitemOpen
  \bibfield  {author} {\bibinfo {author} {\bibfnamefont {R.}~\bibnamefont
  {Casalbuoni}}\ and\ \bibinfo {author} {\bibfnamefont {G.}~\bibnamefont
  {Nardulli}},\ }\href {\doibase 10.1103/RevModPhys.76.263} {\bibfield
  {journal} {\bibinfo  {journal} {Rev. Mod. Phys.}\ }\textbf {\bibinfo {volume}
  {76}},\ \bibinfo {pages} {263} (\bibinfo {year} {2004})},\ \Eprint
  {http://arxiv.org/abs/hep-ph/0305069} {arXiv:hep-ph/0305069} \BibitemShut
  {NoStop}%
\bibitem [{\citenamefont {Anglani}\ \emph {et~al.}(2014)\citenamefont
  {Anglani}, \citenamefont {Casalbuoni}, \citenamefont {Ciminale},
  \citenamefont {Ippolito}, \citenamefont {Gatto}, \citenamefont {Mannarelli},\
  and\ \citenamefont {Ruggieri}}]{Anglani:2013gfu}%
  \BibitemOpen
  \bibfield  {author} {\bibinfo {author} {\bibfnamefont {R.}~\bibnamefont
  {Anglani}}, \bibinfo {author} {\bibfnamefont {R.}~\bibnamefont {Casalbuoni}},
  \bibinfo {author} {\bibfnamefont {M.}~\bibnamefont {Ciminale}}, \bibinfo
  {author} {\bibfnamefont {N.}~\bibnamefont {Ippolito}}, \bibinfo {author}
  {\bibfnamefont {R.}~\bibnamefont {Gatto}}, \bibinfo {author} {\bibfnamefont
  {M.}~\bibnamefont {Mannarelli}}, \ and\ \bibinfo {author} {\bibfnamefont
  {M.}~\bibnamefont {Ruggieri}},\ }\href {\doibase 10.1103/RevModPhys.86.509}
  {\bibfield  {journal} {\bibinfo  {journal} {Rev. Mod. Phys.}\ }\textbf
  {\bibinfo {volume} {86}},\ \bibinfo {pages} {509} (\bibinfo {year} {2014})},\
  \Eprint {http://arxiv.org/abs/1302.4264} {arXiv:1302.4264 [hep-ph]}
  \BibitemShut {NoStop}%
\bibitem [{\citenamefont {Steinhardt}(2019)}]{Steinhardt}%
  \BibitemOpen
  \bibfield  {author} {\bibinfo {author} {\bibfnamefont {P.}~\bibnamefont
  {Steinhardt}},\ }\href@noop {} {\emph {\bibinfo {title} {The Second Kind of
  Impossible: The Extraordinary Quest for a New Form of Matter}}}\ (\bibinfo
  {publisher} {Simon $\&$ Schuster},\ \bibinfo {year} {2019})\BibitemShut
  {NoStop}%
\bibitem [{\citenamefont {Janssen}(1988)}]{JANSSEN198855}%
  \BibitemOpen
  \bibfield  {author} {\bibinfo {author} {\bibfnamefont {T.}~\bibnamefont
  {Janssen}},\ }\href {\doibase https://doi.org/10.1016/0370-1573(88)90017-8}
  {\bibfield  {journal} {\bibinfo  {journal} {Physics Reports}\ }\textbf
  {\bibinfo {volume} {168}},\ \bibinfo {pages} {55} (\bibinfo {year}
  {1988})}\BibitemShut {NoStop}%
\bibitem [{\citenamefont {DiVincenzo}\ and\ \citenamefont
  {Steinhardt}(1999)}]{DiVincenzo}%
  \BibitemOpen
  \bibfield  {author} {\bibinfo {author} {\bibfnamefont {D.}~\bibnamefont
  {DiVincenzo}}\ and\ \bibinfo {author} {\bibfnamefont {P.}~\bibnamefont
  {Steinhardt}},\ }\href@noop {} {\emph {\bibinfo {title} {Quasicrystals: The
  State of the Art}}},\ Series on directions in condensed matter physics\
  (\bibinfo  {publisher} {World Scientific},\ \bibinfo {year}
  {1999})\BibitemShut {NoStop}%
\bibitem [{\citenamefont {Janot}(1997)}]{Janot}%
  \BibitemOpen
  \bibfield  {author} {\bibinfo {author} {\bibfnamefont {C.}~\bibnamefont
  {Janot}},\ }\href@noop {} {\emph {\bibinfo {title} {Quasicrystals: A
  Primer}}},\ Monographs on the physics and chemistry of materials\ (\bibinfo
  {publisher} {Clarendon Press},\ \bibinfo {year} {1997})\BibitemShut {NoStop}%
\bibitem [{\citenamefont {Janssen}\ \emph {et~al.}(2007)\citenamefont
  {Janssen}, \citenamefont {Chapuis},\ and\ \citenamefont
  {et~al.}}]{Jansen:2007}%
  \BibitemOpen
  \bibfield  {author} {\bibinfo {author} {\bibfnamefont {T.}~\bibnamefont
  {Janssen}}, \bibinfo {author} {\bibfnamefont {G.}~\bibnamefont {Chapuis}}, \
  and\ \bibinfo {author} {\bibfnamefont {M.~D.~B.}\ \bibnamefont {et~al.}},\
  }\href@noop {} {\emph {\bibinfo {title} {Aperiodic crystals: from modulated
  phases to quasicrystals}}},\ Vol.~\bibinfo {volume} {20}\ (\bibinfo
  {publisher} {Oxford University Press},\ \bibinfo {year} {2007})\BibitemShut
  {NoStop}%
\bibitem [{\citenamefont {Stadnik}(2012)}]{Stadnik}%
  \BibitemOpen
  \bibfield  {author} {\bibinfo {author} {\bibfnamefont {Z.}~\bibnamefont
  {Stadnik}},\ }\href@noop {} {\emph {\bibinfo {title} {Physical Properties of
  Quasicrystals}}},\ Springer Series in Solid-State Sciences\ (\bibinfo
  {publisher} {Springer Berlin Heidelberg},\ \bibinfo {year}
  {2012})\BibitemShut {NoStop}%
\bibitem [{\citenamefont {Fan}(2016)}]{Fan}%
  \BibitemOpen
  \bibfield  {author} {\bibinfo {author} {\bibfnamefont {T.}~\bibnamefont
  {Fan}},\ }\href@noop {} {\emph {\bibinfo {title} {Mathematical Theory of
  Elasticity of Quasicrystals and Its Applications}}},\ Springer Series in
  Materials Science\ (\bibinfo  {publisher} {Springer Singapore},\ \bibinfo
  {year} {2016})\BibitemShut {NoStop}%
\bibitem [{\citenamefont {Scott}\ and\ \citenamefont {Clark}(2012)}]{Scott}%
  \BibitemOpen
  \bibfield  {author} {\bibinfo {author} {\bibfnamefont {J.}~\bibnamefont
  {Scott}}\ and\ \bibinfo {author} {\bibfnamefont {N.}~\bibnamefont {Clark}},\
  }\href@noop {} {\emph {\bibinfo {title} {Incommensurate Crystals, Liquid
  Crystals, and Quasi-Crystals}}},\ Nato Science Series B\ (\bibinfo
  {publisher} {Springer US},\ \bibinfo {year} {2012})\BibitemShut {NoStop}%
\bibitem [{\citenamefont {Jaric}\ \emph {et~al.}(1988)\citenamefont {Jaric},
  \citenamefont {Jaric}, \citenamefont {Bak},\ and\ \citenamefont
  {Gratias}}]{Jaric}%
  \BibitemOpen
  \bibfield  {author} {\bibinfo {author} {\bibfnamefont {M.}~\bibnamefont
  {Jaric}}, \bibinfo {author} {\bibfnamefont {M.}~\bibnamefont {Jaric}},
  \bibinfo {author} {\bibfnamefont {P.}~\bibnamefont {Bak}}, \ and\ \bibinfo
  {author} {\bibfnamefont {D.}~\bibnamefont {Gratias}},\ }\href@noop {} {\emph
  {\bibinfo {title} {Introduction to Quasicrystals}}},\ Advances in Veterinary
  Medicine\ (\bibinfo  {publisher} {Academic Press},\ \bibinfo {year}
  {1988})\BibitemShut {NoStop}%
\bibitem [{\citenamefont {Baggioli}\ and\ \citenamefont
  {Landry}(2020)}]{Baggioli:2020haa}%
  \BibitemOpen
  \bibfield  {author} {\bibinfo {author} {\bibfnamefont {M.}~\bibnamefont
  {Baggioli}}\ and\ \bibinfo {author} {\bibfnamefont {M.}~\bibnamefont
  {Landry}},\ }\href {\doibase 10.21468/SciPostPhys.9.5.062} {\bibfield
  {journal} {\bibinfo  {journal} {SciPost Phys.}\ }\textbf {\bibinfo {volume}
  {9}},\ \bibinfo {pages} {062} (\bibinfo {year} {2020})},\ \Eprint
  {http://arxiv.org/abs/2008.05339} {arXiv:2008.05339 [hep-th]} \BibitemShut
  {NoStop}%
\bibitem [{\citenamefont {Sur\'owka}(2021)}]{Surowka:2021ved}%
  \BibitemOpen
  \bibfield  {author} {\bibinfo {author} {\bibfnamefont {P.}~\bibnamefont
  {Sur\'owka}},\ }\href {\doibase 10.1103/PhysRevB.103.L201119} {\bibfield
  {journal} {\bibinfo  {journal} {Phys. Rev. B}\ }\textbf {\bibinfo {volume}
  {103}},\ \bibinfo {pages} {L201119} (\bibinfo {year} {2021})},\ \Eprint
  {http://arxiv.org/abs/2101.12234} {arXiv:2101.12234 [cond-mat.str-el]}
  \BibitemShut {NoStop}%
\end{thebibliography}

%

\end{document}